\documentclass[pre,floatfix,superscriptaddress,twocolumn]{revtex4-1}
\usepackage{graphicx}
\usepackage{subfigure}
\usepackage{bm}
\usepackage{amsmath}
\usepackage{lineno}
\usepackage{amssymb}
\usepackage{lipsum}
\usepackage{appendix}
\usepackage{color}
\usepackage{float}

\begin{document}

\title{The conformation of a semiflexible filament in a quenched random potential}

\author{Valentin M. Slepukhin}
\affiliation{Department of Physics and Astronomy, UCLA, Los Angeles California, 90095-1596, USA}

\author{Maximilian J. Grill}
\affiliation{Technical University of Munich, Germany, Department of Mechanical Engineering, Institute for Computational Mechanics}

\author{Kei W. M\"{u}ller}
\affiliation{Structural and Applied Mechanics Group, Computational Engineering Division, Lawrence Livermore National Laboratory,
7000 East Ave, Livermore, California 94550, USA}
\affiliation{Technical University of Munich, Germany, Department of Mechanical Engineering, Institute for Computational Mechanics}

\author{Wolfgang A. Wall}
\affiliation{Technical University of Munich, Germany, Department of Mechanical Engineering, Institute for Computational Mechanics}

\author{Alex J. Levine}
\affiliation{Department of Physics and Astronomy, UCLA, Los Angeles California, 90095-1596, USA}
\affiliation{Department of Chemistry and Biochemistry, UCLA, Los Angeles California, 90095-1596, USA}
\affiliation{Department of Biomathematics, UCLA, Los Angeles California, 90095-1596, USA}

\begin{abstract}

Motivated by the observation of the storage of excess elastic free energy -- {\em prestress} -- in cross linked semiflexible networks, we consider the
problem of the conformational statistics of a single semiflexible polymer in a quenched random potential. The random potential, which represents
the effect of cross linking to other filaments is assumed to have a finite correlation length $\xi$ and mean strength $V_{0}$.  We examine
statistical distribution of curvature in filament with thermal persistence length $\ell_{P}$ and length $L_0$ in the limit that $\ell_{P} \gg L_0$.  We
compare our theoretical predictions to finite element Brownian dynamics simulations. Lastly we comment on the validity of replica field techniques in
addressing these questions.
\end{abstract}

\pacs{XXX}

\date{\today}

\maketitle

\section{Introduction}
\label{sec:Introduction}

Semiflexible polymer networks are
well known to trap {\em prestress} in their formation.  Cross linking molecules typically lock-in curved and thus
elastically stressed states of the filaments.  As a consequence of this being an out of equilibrium process, the cross linkers may, however,
trap more thermal energy -- $k_{\rm B}T$ -- per bending mode.  The result is that cross-linking during network formation
typically traps excess free energy, which then slowly bleeds out of the system.  That relaxation process appears to
lead to large, nonequilibrium stress fluctuations, and is associated with
the glassy power-law rheology of the network at very low frequencies. Such low-frequency power-law rheology has been
observed in both simulation~\cite{Levine:14} and experiment~\cite{Broedersz:10}. Living cells are similarly observed to have soft, glassy, power-law
rheology~\cite{Sollich:97}, albeit with a distinctly different power law exponent~\cite{Lenormand:04,Fredberg:05}.
The large nonequilibrium stress fluctuations have currently
been observed solely in simulation and we suggest that they should make an intriguing target for future experiments.

The underlying dynamics of transiently cross linked
semiflexible networks is likely to be fundamental to the mechanics of both the active biological and passive {\em in vitro} systems.
Simulations suggest that both the stress fluctuations and this
characteristic power-law rheology of transiently cross linked networks are associated with the reorganization of progressively larger
sections of the network occurring on progressively
longer time scales. Understanding these dynamics presents a theoretical challenge.

In an effort to better understand the excess free energy
trapped in such networks, we consider the problem of a single semiflexible filament at temperature
$T$ interacting with a quenched random potential.  We hope that, in this single filament model, the quenched pinning potential mimics the
effect of random cross linking sites that mechanically couple the filament in question to the surrounding network.
In particular, we will examine the role of
spatial correlations in the pinning potential, which, at least loosely speaking, introduces an effective mesh size of the surrounding network.
We analyze the disordered-averaged conformational fluctuations of the filament and the elastic energy stored in the system as
a function of persistence length of the filament and the correlation length of the potential. We compare these predictions to the results
of large-scale finite element Brownian dynamics simulations of such a semiflexible filament in a random potential.

The study of the statistical mechanics of a single stiff filament in a random potential recalls a number of related systems in which a
low-dimensional elastic object interacts with a quenched pinning potential. Examples include disorder-pinned
domain walls between symmetry-equivalent ground
states~\cite{Kuntz:00}, vortex lines in superconductors~\cite{Marchetti:90,Hwa:93,Ertas:94}, and
the three-phase contact line associated with the spreading of fluids as they wet a disordered substrate~\cite{Narayan:93,Schaffer:00}.
The key distinction between
these systems and the one of current interest is the presence of a bending term in the elastic Hamiltonian of the filament.  In cases where
this term dominates the statistical weights of various filament conformations, {\em i.e.}, when filament tension is sufficiently small or when
examining bends on short enough lengths, we expect to obtain results distinct from those obtained for these previously studied systems.

A filament interacting with a quenched random potential can be characterized by three lengths: the filament length $L_0$, the thermal
persistence length $\ell_{\rm P}$, and the correlation length $\xi$ of the potential.
We will focus on the case of stiff filaments (although we introduce alternatives) in which the persistence length is typically longer than that of the
filament itself: $L_0 < \ell_{\rm P}$.  In this case, there are still two distinct limits.  One might imagine that either $\ell_{\rm P} \ll \xi$, in which case the
filament should be flexible enough to follow the twists and turns of the local minima of the random potential. Alternatively one may consider
the case where $\ell_{\rm P} \gg \xi$ and the filament is so stiff that the elastic energy cost for following the
valleys of the potential becomes prohibitive.

Similarly, the problem is endowed with two energy scales: the thermal energy $T$ (we work in units where Boltzmann's
constant is set equal to unity)  and the typical energy scale of the pinning potential $V_{0}$. The potential has
dimensions of energy per length.   The inverse length
scale $ \nu = V_{0}/T$ must control the states of the filament, allowing one to examine both ``strongly pinned'' on scales where
$L_0 \nu \gg 1$ versus ``weakly pinned'' $L_0 \nu \ll 1$ cases.  The former is more interesting  for the system under consideration.

There are two quantities that will provide insight into the ensemble of filament configurations.  These are the effective persistence length of
a filament in the pinning potential. This length differs from the usual tangent vector correlation length of the filament due to its interaction
with the pinning potential. It also provides a clear target for experimental studies of labeled filaments in networks.  The effect of this extra
filament bending imposed by the pinning potential (or the network in experiment) is that the filaments store excess elastic energy. We
propose that this excess free energy is a prediction for the prestress in networks. The single filament model thus makes two predictions for
nonequilibrium networks.

The remainder of this article is organized as follows. In section~\ref{sec:model} we describe the filament Hamiltonian.  We then provide
in section~\ref{sec:Strong-pinning} an analytical calculation of the averaged free energy of the filament in the strong-pinning
limit of a random potential,
described above. We start with the case of the completely flexible filament, i.e., the filament without bending energy,
and then move forward to the more general case of the semiflexible filament. We then turn in section~\ref{sec:simulations} to
numerical simulations of the system using finite element Brownian dynamics of a geometrically exact Simo-Reissner beam model. We then summarize
our results and their implications for pre-stress in networks in section~\ref{sec:discussion}, where we conclude with a
proposal for new experiments. Finally, we provide an appendix addressing the applicability of the replica method to the problem explaining that
it provides unphysical results for many of the measurable quantities.

\section{The model}
\label{sec:model}

\begin{figure}
\includegraphics[width=8cm]{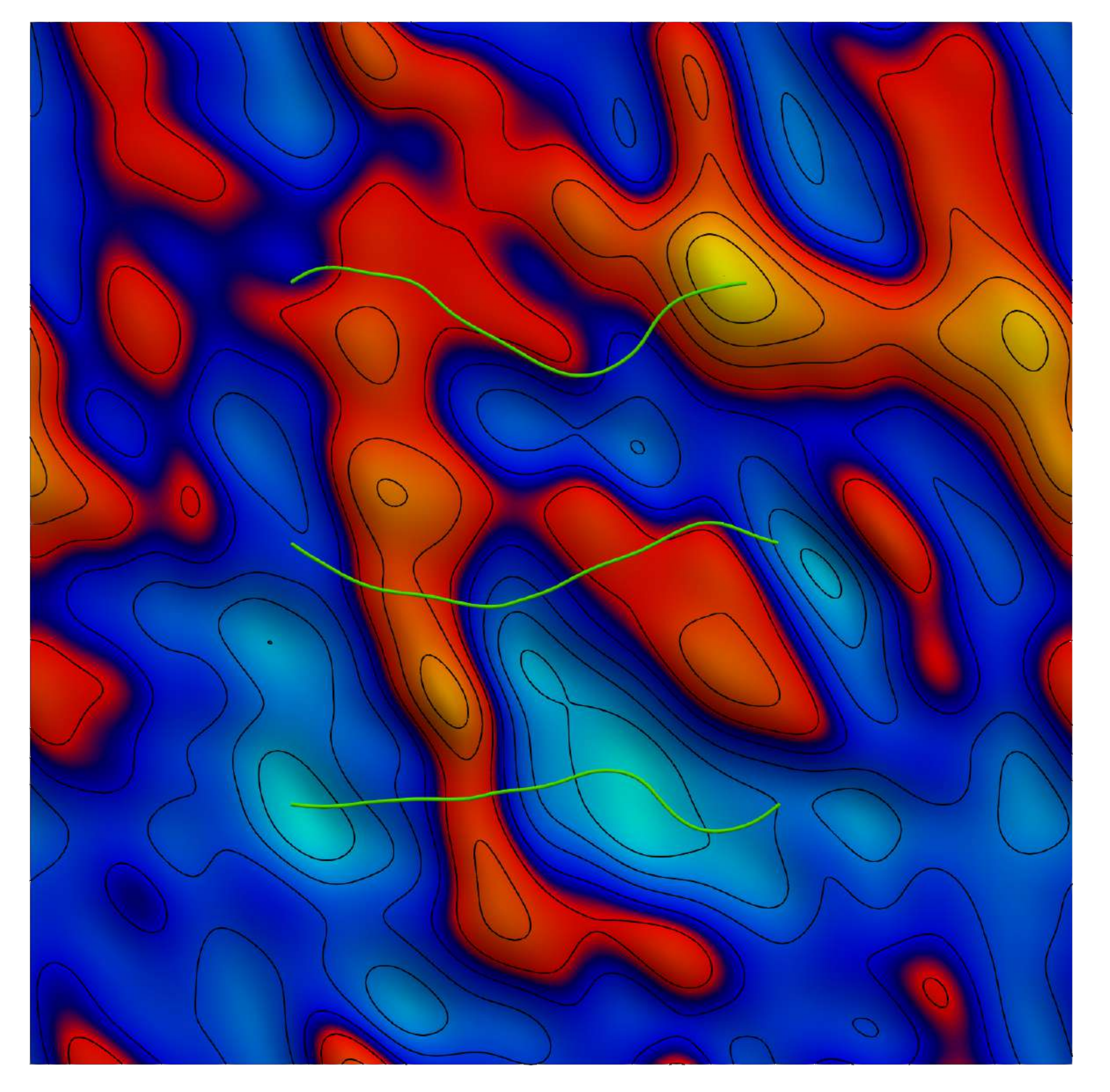}
\caption{(color online) Three tensed  filaments interacting with the pinning potential, shown as a heat map with hotter colors representing higher potential energies.
The lowest filament traverses a saddle between local
potential maxima.  On the right  of that saddle point it curves into a deep potential minimum (blue). Similar features
may be seen in the other filaments. This is a snapshot from our Brownian dynamics simulations, discussed in section~\ref{sec:simulations}.}
\label{fig:schematic}
\end{figure}

We consider a filament in two dimensions that is anchored at both ends.  We introduce a coordinate system in which these anchoring
sites are at positions $(0,0)$ and  $(L_0,0)$ respectively. The anchoring sites are assumed to be able to generate arbitrary
constraint forces necessary to hold the filament at this points, but to provide no constraint torques.  The directed filament is
further assumed to be free of overhangs, allowing its configuration to be described by a function $y(x)$.  A
representation for a numerical simulation (to be described in section~\ref{sec:simulations}) of
a few filaments (green lines) interacting with the potential (heat map)  is shown in Fig.~\ref{fig:schematic}.  We will allow the arclength of the
filament between these two anchoring points to vary.  In short, the anchoring points are reservoirs of extra length.
A simple mechanical model of this situation can be thought of as follows. The filament fluctuates on a table whose
height topography in a uniform gravitation field gives the pinning potential.  The anchoring points may be thought of as holes in this table through
which more filament may enter or exit the table's surface. Weights may also be added below the table to enforce a fixed tension on the filament.
We should notice that for the simulation we use a slightly different model, with one end fixed and the other free to move in one direction (see section \ref{sec:simulations}). We assume that for the small conformation the difference between these two models is negligible.

In the small bending limit ($\frac{dy}{dx} \ll 1$), which should be valid for filaments much shorter than their persistence length,  we can
write the filament's energy functional as
\begin{equation}
\label{Energy}
E[y(x)] = \int_0^{L_0} dx \left\{ \frac{1}{2} \kappa \ddot{y}^2 + \frac{1}{2} \tau \dot{y}^2 + V \left( x ,y \left(x \right) \right) \right\},
\end{equation}
where $\kappa$ is the filament's bending modulus defining a thermal persistence length $\ell_{\rm P} = \kappa/T$.   $\tau$ is the tension
imposed on the filament.  $V(x,y)$ is the quenched random potential (with dimensions of energy per length),
described in more detail below. We have introduced the notation
$\dot{y} = \frac{d y}{d x}$.  Then the classical partition function for such a filament at temperature $T = 1/\beta$ is given by the path
integral over all trajectories of the filament weighted by a Boltzmann factor obtained from Eq.~\ref{Energy}:
\begin{equation}
\label{Partition-Function}
Z  = \int {\cal D} y(x) \, e^{- \beta E[y(x)]}.
\end{equation}
We will later consider averages of the free energy, obtained from Eq.~\ref{Partition-Function} in the usual way: $F = - T \ln Z$, over an
ensemble of random potentials.

Turning to the pinning potential, we can consider two rather simple forms for its probability distribution that allow for a finite spatial
correlation length, but do not break rotational invariance.   The first is inspired by the massive scalar field Lagrangian
\begin{equation}
\label{potential-distribution-1}
{\cal P}_{\rm V}(V) = \frac{1}{P_{0}} \exp \left\{ - \frac{1}{2 V_{0}^{2}} \int d^{2} x  \left[ (\nabla V)^2 +  \xi^{-2} V^2 \right]  \right\}.
\end{equation}
Here $\xi$ sets the correlation length and $V_{0}$ the energy scale of the potential.  It is straightforward to see that this model generates
an ensemble of random potentials in which the amplitude of
each Fourier mode is selected as an independent Gaussian random variable from a distribution with zero mean and a width
that depends on the magnitude of the wavenumber $k = | {\bf k} | $.

We may also consider a related problem in which the correlation length is assigned to the force rather than to the potential. Since the
force is the (negative) derivative of the potential, one can obtain the necessary Gaussian probability distribution for force by introducing
another derivative in Eq.~\ref{potential-distribution-1}.  We obtain
\begin{equation}
\label{Force-Distribution}
{\cal P}_{\rm F}(V) \sim  \exp \left[ - \frac{1}{2 V_{0}^{2}} \int d^{2}x \,  \xi^2 ( \nabla^{2} V)^2 + (\nabla V)^2 \right].
\end{equation}

The first version of the potential in which the Fourier amplitudes of the scalar potential are Gaussian distributed generates quite large
pinning forces at short correlation lengths since the slope of the potential is $V_{0}/\xi$.
We speculate that the second version of the potential in which the
spectrum of {\em pinning forces} on the filament is Gaussian distributed is a better approximation of the physical problem since one may then
manipulate the correlation length of the potential (which is our approximation to the mesh size of a filament network) without changing the
scale of forces to which the filament is subjected. We return to this point in our discussion. Hereafter we refer to the first type of random
potential as the {\em energy controlled} distribution while the second will be called the {\em force controlled} distribution.

\begin{figure}[htb]
\subfigure[Energy controlled distribution.]
{
  \includegraphics[width=6cm]{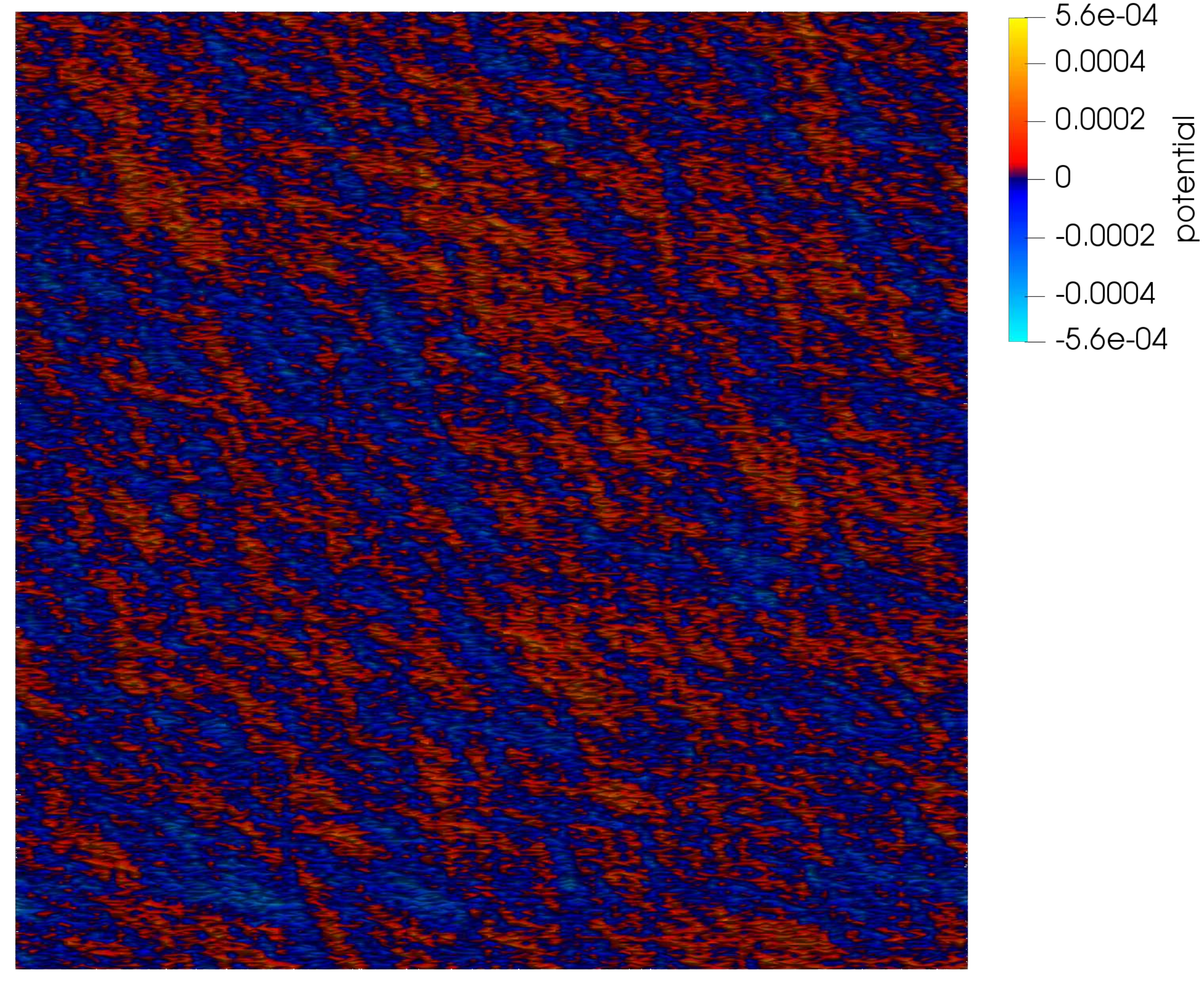}
  \label{fig:pot_field_powerlaw_quadratic}
}
\subfigure[Force controlled distribution.]
{
  \includegraphics[width=6cm]{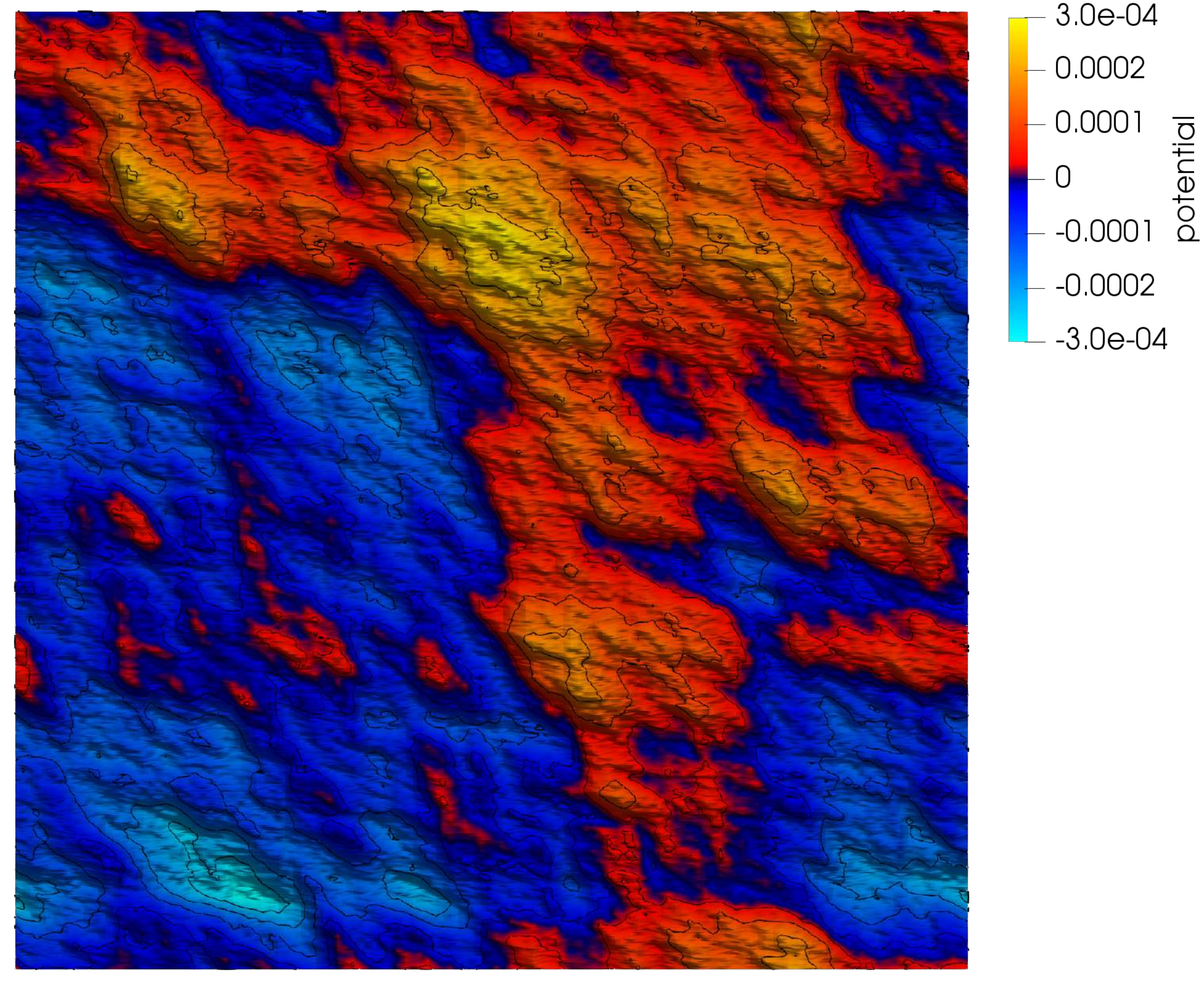}
  \label{fig:pot_field_powerlaw_quartic}
}
\subfigure[Exponential suppression of high modes.]
{
  \includegraphics[width=6cm]{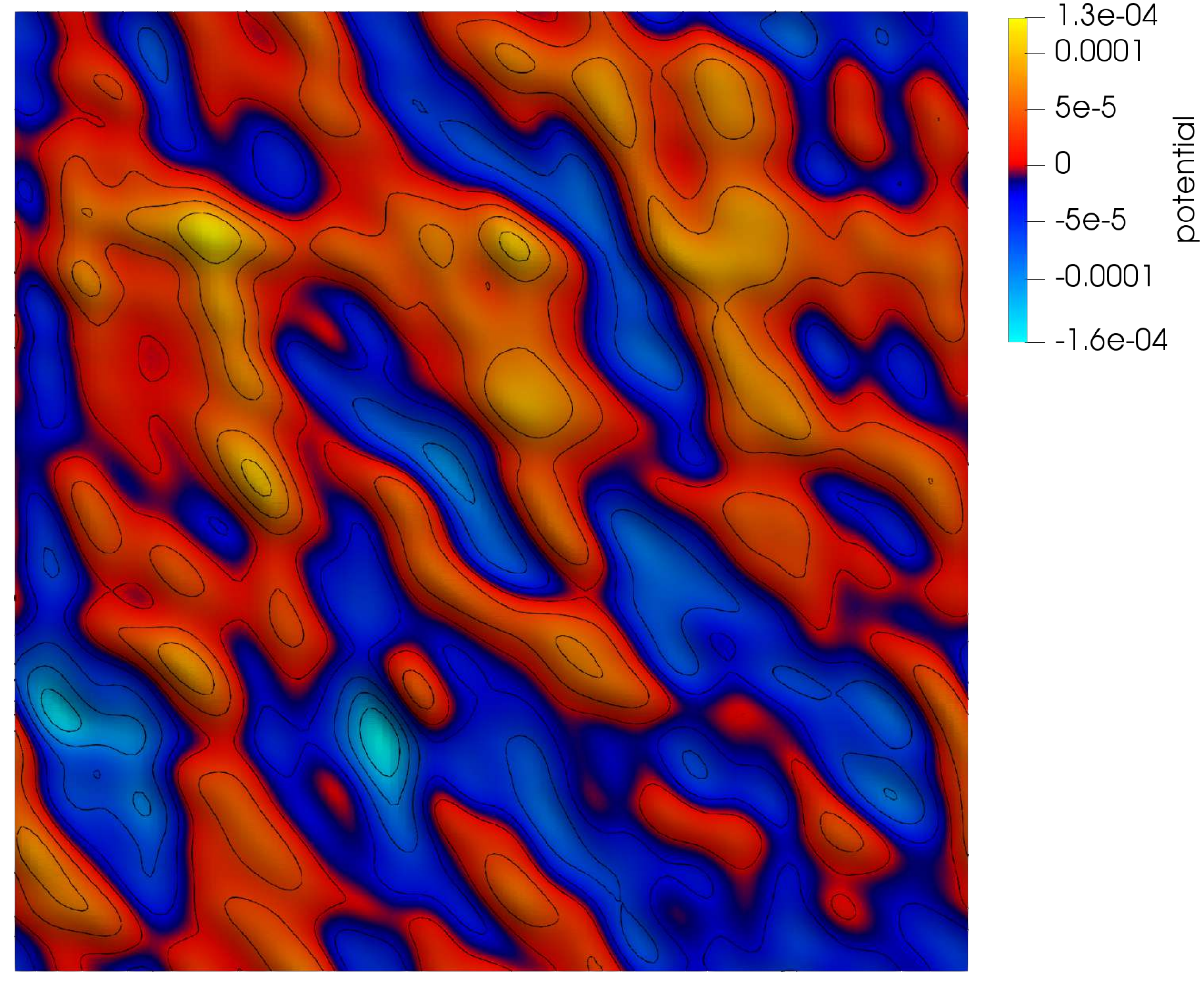}
  \label{fig:pot_field_exponential}
}
\caption{(color online) Examples of the random potentials~$V(x,y)$ (shown as a heat map with contour lines) selected from different distributions, labeled in the figure.
The correlation length is fixed in all three so that $L_{x}/\xi = 20$ and the energy scale is $V_0=10^{-4}$.}
\label{fig:pot_field_comparison}
\end{figure}

\begin{figure}[htb]
\subfigure[Force controlled distribution.]
{
  \includegraphics[width=6cm]{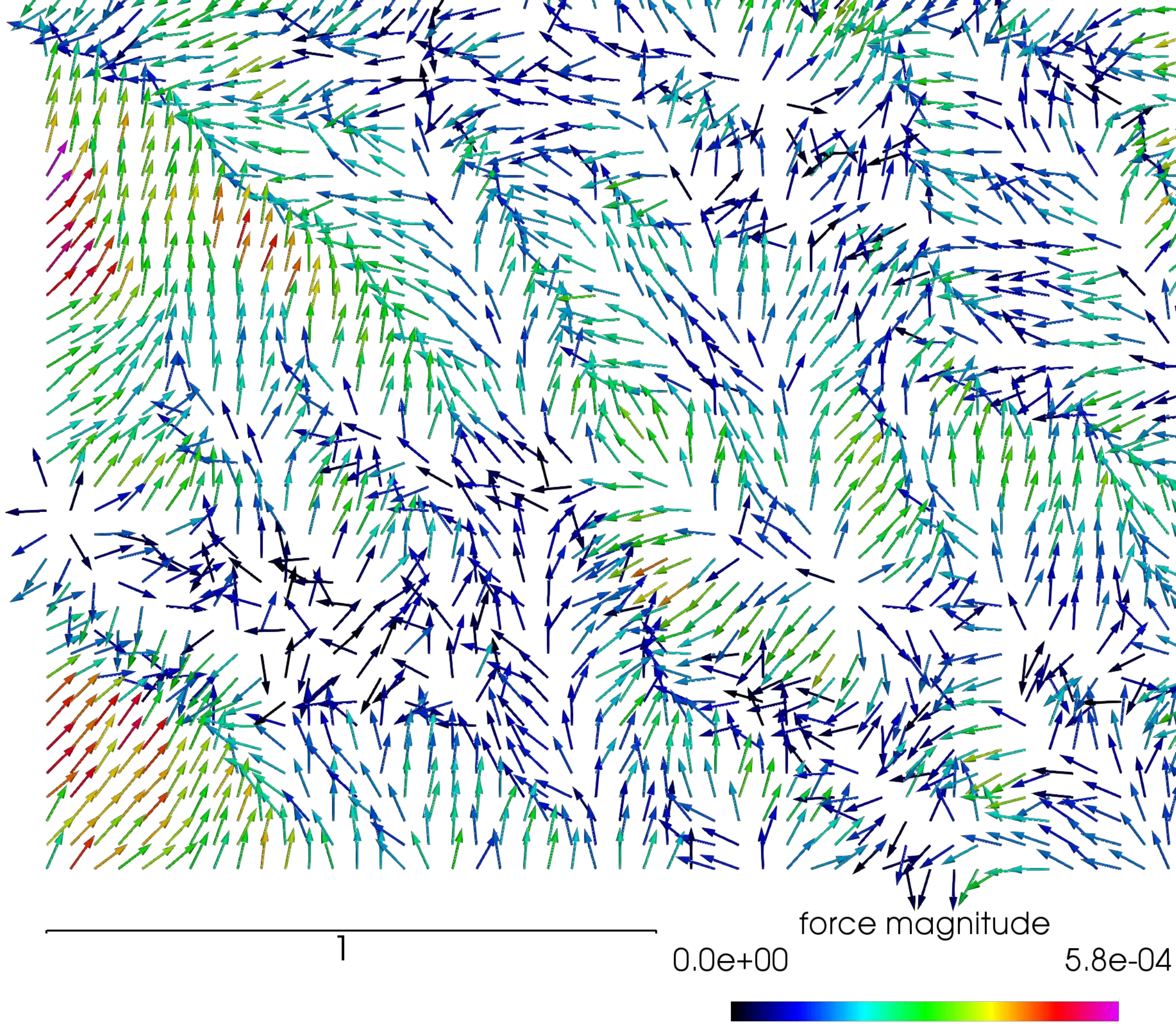}
  \label{fig:force_field_quartic}
}
\subfigure[Exponential suppression of high modes.]
{
  \includegraphics[width=6cm]{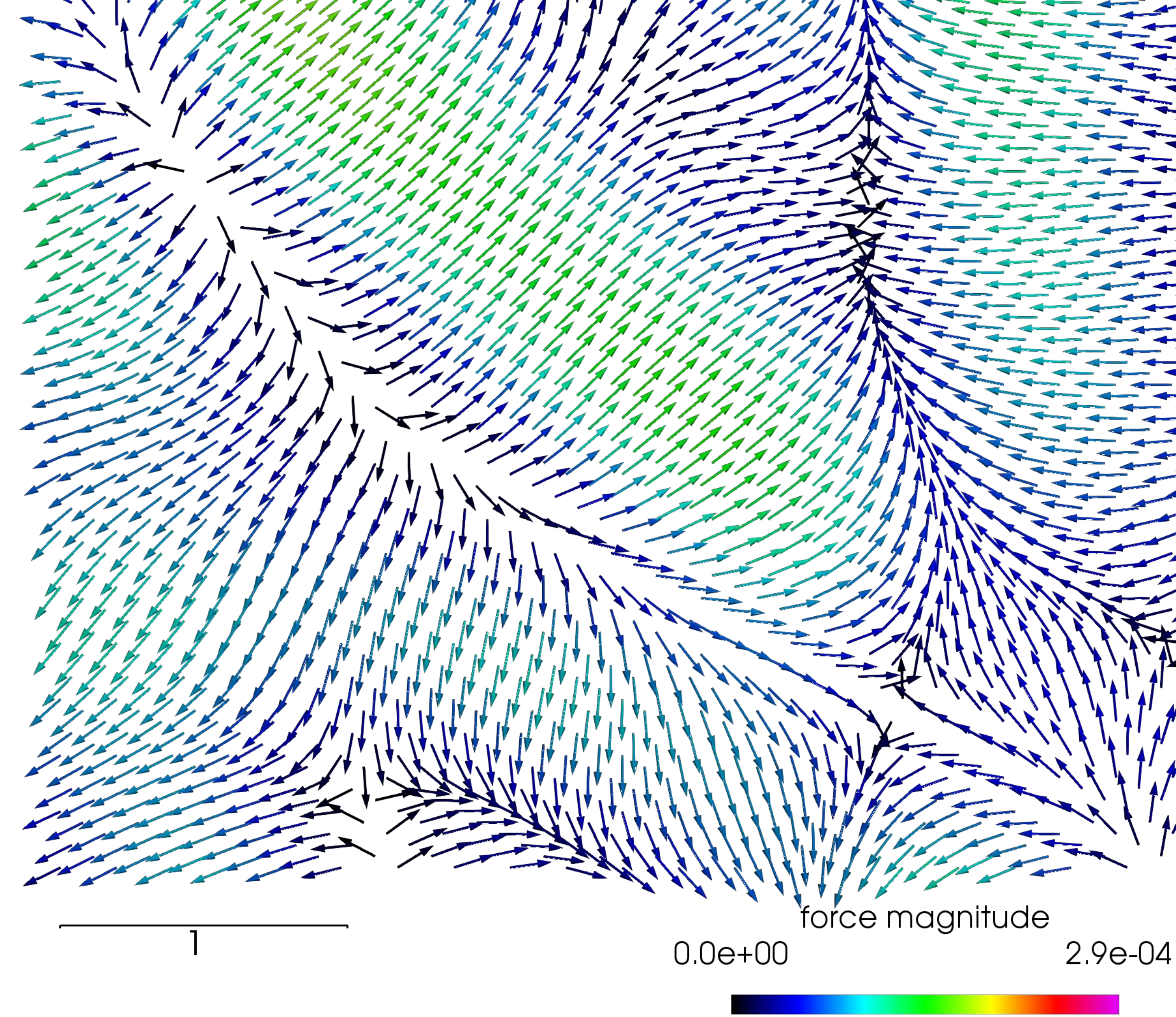}
  \label{fig:force_field_exponential}
}
\caption{(color online) Force fields resulting due to the random pinning potentials shown in Fig.~\ref{fig:pot_field_comparison}.}
\label{fig:force_field_comparison}
\end{figure}

These distributions can also be expressed in terms of a probability distribution for the Fourier components of the potentials.
Working in terms of those Fourier modes, we can also introduce another distribution with an exponential suppression of the higher Fourier modes:
\begin{equation}
\label{Exponential-suppression}
P(V_{k_x,k_y}) \propto \exp \left\{ -\frac{L_x L_y}{\xi^2}  \frac{V_{k_x,k_y}^2}{8 V_{0}^2}  \exp \left(\xi^2 (k_x^2 + k_y^2) \right)  \right\},
\end{equation}
where the rectangular system has an area $A = L_{x} L_{y}$.
This particular form of  the random potential is not convenient for analytic calculations, but generates more numerically stable simulations.  See
Fig.~\ref{fig:pot_field_comparison} for examples of random potentials selected from these distributions.  The pinning forces generated from these
potentials are shown in Fig.~\ref{fig:force_field_comparison}. The energy controlled potential produces a complex force landscape with very short ranged correlations.
We do not reproduce that vector field here.

\section{The Valley Approximation}
\label{sec:Strong-pinning}
There is a considerable simplification to be found if we assume that $ L \nu \gg 1$. In this strong pinning limit, the ensemble of filament
configurations is dominated by states where the filament is confined to
the valleys of the pinning potential.  We indeed observe this in simulations.  There we also see instances in which filament crosses from
one valley to another over a saddle point of the potential. We will address escapes
from one valley to the next over saddle points later.  Excluding such saddle points for now, we assume that the
potential is roughly constant along the bottom  of the valley and that the
curvature of the potential in the direction orthogonal to the path along the valley floor is also constant.  Thus, the local
form of the pinning potential is given by
\begin{equation}
\label{harmonic-approximation-potential}
 V(x,y) =\frac{\tilde{k}}{2} \left[ y - y_0 (x) \right]^2.
\end{equation}
We have introduced a curvature (spring constant) scale $\tilde{k} =  \frac{V_0}{ \xi^2} $.  The path of the
valley minimum $y_0 (x) $ remains a random curve.  To analyze the effect of the quenched distribution of such paths, we may either calculate
physical quantities of interest for an arbitrary curve $y_0(x)$ and then average, or use the replica trick to handle the average over the potential
simultaneously with the thermal averaging.  The replica trick, however, provides unphysical results as we explain in appendix C.

\subsection{Flexible filaments with the energy controlled distribution}
We now explore the valley approximation first for a  flexible polymer by setting $\kappa =0$.  We require a finite tension so the polymer's
path can be considered to be nearly straight.  In this limit the energy of the polymer in the random potential may be written as
\begin{equation}
\label{flexible-energy}
\frac{E_{\rm flexible}}{T} = \int_0^{L_0} dx \left\{  y(x) \mathcal{O} y(x) +  k y(x) y_0(x)  + \frac{k y_0^2 (x)}{2} \right\},
\end{equation}
where  we have scaled the parameters by temperature $T$: $m = \beta \tau$ and $k = \beta \tilde{k}$.  We have also
introduced the differential operator
\begin{equation}
\label{flexible-operator}
\mathcal{O} = \frac{m}{2} \partial^2 + \frac{k}{2}.
\end{equation}
The partition function is given by the integral
\begin{equation}
\label{flexible-partition-function}
Z_{\rm flexible} = \int {\cal D}y e^{- E_{\rm flexible}[y(x)]}.
\end{equation}

Calculating this partition sum is equivalent to performing the Euclidean path integral (with $x$ being the time-like coordinate)
for a quantum particle with a mass $m$ in a harmonic potential with
spring constant $k$.  In the analogous quantum problem the effect of the pinning potential is to introduce
a time-dependent force $-k y_0(x)$. Since the integral is Gaussian, we can obtain
a closed form solution for the partition function for a particular path of the valley floor:
\begin{equation}
\label{flexible-partition-1}
Z =\frac{1}{\sqrt{\det \mathcal{O}}} \exp \left[ -\int_0^{L_0} dx \left( -  \frac{k^2}{4} y_0(x) \mathcal{O}^{-1} y_0(x)  + \frac{k  y_0^2}{2} \right) \right].
\end{equation}
If we choose $y_0(x) = 0$ everywhere, Eq.~\ref{flexible-partition-1} reduces to the partition function of an unforced oscillator.
From this we obtain the prefactor, leading us to write
\begin{equation}
\label{flexible-partition-2}
\frac{Z}{Z_{\rm SHO}} =  \exp \left[ -\int_0^{L_0} dx \left( -  \frac{k^2}{4} y_0(x) \mathcal{O}^{-1} y_0(x)  + \frac{k  y_0(x)^2}{2} \right) \right],
\end{equation}
where $Z_{\rm SHO}$ is the well-known result for the simple harmonic oscillator~\cite{Kleinert-path-integrals}. It is then straightforward to compute the
free energy $F$ of the flexible chain in a particular realization of the random potential, the one whose valley follows the path $y_0(x)$.  This
free energy is given by
\begin{equation}
\label{Free-energy-flexible}
F =F_{S H O}  + T \int_0^{L_0} dx \left( -  \frac{k^2}{4} y_0(x) \mathcal{O}^{-1} y_0(x)  + \frac{k  y_0(x)^2}{2} \right).
\end{equation}
We now average the free energy in Eq.~\ref{Free-energy-flexible} over a distribution of the paths of the valley floor that is consistent with
our previous random potential distribution given by Eq.~\ref{potential-distribution-1}.
To obtain this we weight the paths of the valley floor $y_0(x)$ by
\begin{equation}
\label{xb-distribution}
{\cal P} \left[y_0(x) \right] \sim \exp \left[ - \frac{1}{2 \xi }  \int_0^{L_0} d x \dot{y}_b^2  \right] = e^{-\int_0^{L_0} y_0 \mathcal{G} y_0 dx},
\end{equation}
where differential operator
\begin{equation}
\label{valley-path-weight}
\mathcal{G}  = -\frac{1}{2 \xi} \partial^2
\end{equation}
incorporates the correlation length of the pinning potential $\xi$.
This weight is analogous to the Euclidean path integral of a free particle.  The distribution of quenched potentials
determined by Eqs.~\ref{harmonic-approximation-potential} and \ref{xb-distribution} is not identical to that given by
Eq.~\ref{potential-distribution-1}.  But the statistical weight associated with valley floors of these potentials has the same
spatial correlations as those valley floors determined by the
original potential distribution in Eq.~\ref{potential-distribution-1}.

The averaging over the distribution of these valley floors we obtain a correction to the simple harmonic oscillator (SHO) free energy
\begin{equation}
\left[ F \right]   = F_{S H O} + \Delta F,
\end{equation}
where
\begin{equation}
\label{free-energy-change}
\Delta F =  -  \frac{T k^2}{8} \mbox{Tr} ( \mathcal{O}^{-1} G^{-1} )    + \frac{  Tk  }{4} \mbox{Tr} G^{-1}.
\end{equation}
Here and throughout this article, we use the squared brackets $\left[ \cdot \right]$ to indicate averages with respect to the
quenched random potential.  The angled brackets $\langle \cdot \rangle$ represent thermal averages. We
compute these traces by diagonalizing the two relevant operators -- see appendix~\ref{app:traces}.  We find that for
long filaments (see appendix A)
the disorder-averaged free energy is
\begin{equation}
\label{flexible-filament-free-energy}
\left[ F \right]  = \frac{L_{0}  \sqrt{V_{0} \tau}}{4} \left[ 1 +  \frac{2 T}{\xi \tau} \right].
\end{equation}

The correlation length of the quenched potential and the tension set a natural energy scale that controls the free energy correction.
When that potential is sufficiently heterogeneous so that its valleys are
quite tortuous on the scale of a Pincus blob~\cite{Pincus:76} $ \xi < T/\tau$, the pinning potential has a significant effect on the
free energy.  The result above -- Eq.~\ref{flexible-filament-free-energy} -- cannot be extended to arbitrarily
small tensions $\tau \rightarrow 0$ since our assumption that $x(y)$ is a well-defined function breaks down. In effect our use of the Monge gauge
fails to adequately represent the polymer's shape.

Finally, we compute the contribution to the length, as compared with the length of the filament in the absence of the pinning potential. We
recall that the contour length of the filament
\begin{equation}
\label{contour-length}
L = \int_{0}^{L_{0}} dx \sqrt{1 + \dot{y}^{2}} \approx L_{0} + \frac{1}{2}\int_{0}^{L_{0}} dx \,  \dot{y}^{2}.
\end{equation}
where the Taylor expansion of the integrand is justified by the fact that the filament is nearly straight when under sufficient tension.
The quantity of interest is thermal
average $\langle L \rangle$.  As the integrand is nonnegative, this average $\langle L \rangle $ is necessarily
longer than the separation of the end points $L_{0}$. In fact,
in the absence of a bending modulus (as  assumed here), $\langle L \rangle $ is divergent; there is an infinite amount of
contour length trapped in the high wavenumber modes of deformation.
It is thus useful to define the change in the contour length of the polymer due to the quenched pinning potential. We introduce
\begin{equation}
\label{definition-excess-length}
\Delta L = \left[ \langle L \rangle \right] - \left. \langle L \rangle\right|_{V_{0}=0}
\end{equation}
as the difference between the thermal average of the filament's contour length when averaged again over ensemble of pinning potentials at
fixed $V_{0}$ and $\xi$ and the same quantity with no pinning potential, {\em i.e.}, the problem obtained by setting $V_{0}=0$.
This difference remains finite. We find that
\begin{equation}
\label{stretching}
\frac{\Delta L}{L_{0}}  =  \frac{1}{8} \sqrt{\frac{V_{0}}{\tau}} \left[ 1 - \frac{2 T}{\tau \xi} \right].
\end{equation}

Once again we see that the corrections due to the pinning potential enter through the dimensionless ratio of the
Pincus blob length to the correlation length of the potential.  The negative sign in Eq.~\ref{stretching} may appear to be counterintuitive.  One might
imagine that a shorter correlation length would, in fact, create more transverse undulations in the filament as it tries to follow the more sinuous
potential minimum. A shorter correlation length $\xi$ would then be expected to increase $\Delta L$. It does not.  The tensed
{\em flexible} polymer has large undulations on length scales below that of the Pincus blob.  The effect of a decreasing potential correlation
length at fixed $V_{0}$ is to increase the curvature of the potential, making the harmonic constraint forces on the filament stronger. These larger
forces straighten out the filament on scales below the Pincus blob length, {\em decreasing} $\Delta L$ by straightening it out on
these small scales.

The first term in Eq.~\ref{stretching} increases the $\Delta L$ of the filament with increasing potential (or decreasing $\tau$).  This reflects the
expected effect of the potential.  In a stronger potential the filament is forced to follow more precisely the tortuous valley of the potential minimum
and thereby use more arclength.  We return to this idea in the case of semiflexible filaments. From this analysis, we see that one
may wish to consider an ensemble of random potentials for which the typical scale of the pinning forces remains fixed even as the
correlation length is changed.

\subsection{Semiflexible filaments: energy controlled distribution}

We now include the bending modulus in the filament Hamiltonian.   Using the same valley-based approximation for the random potential, the
energy of the {\em semiflexible} filament takes the form
\begin{equation}
\label{semiflexible-energy}
\frac{E}{T} =  \int_0^{L_0} dx \left[  \frac{\ell_{\rm P} \ddot{y}^2}{2} + \frac{m \dot{y}^2}{2} + \frac{k (y(x) - y_0(x))^2}{2} \right] .
\end{equation}
We introduce $\ell_{\rm P} = \beta \kappa$, the normal thermal persistence length of the filament.  The partition sum is then given by
\begin{equation}
\label{partition-function-bending-1}
Z =  \int {\cal D}y(x) e^{-E/T}.
\end{equation}
In Eq.~\ref{partition-function-bending-1} we restrict the paths by imposing boundary conditions such that the filament begins and
ends at $x=0,L_{0}$ respectively.  Moreover, it starts and ends at zero tangent angle with respect to the mean direction (along the $x$ axis):
$\dot{y}(0) = \dot{y}(L_{0}) =0$.

Following our previous procedure, we may formally integrate over all paths $y(x)$ by introducing the inverse of the differential operator
\begin{equation}
\label{bending-operator}
\mathcal{O}_{\kappa} = \frac{\ell_{\rm P}}{2} \partial^4 - \frac{m}{2} \partial^2 + \frac{k}{2}
\end{equation}
and write the partition function as
\begin{equation}
Z =\frac{1}{\sqrt{\det \mathcal{O}_{\kappa}}} e^{-\int_0^{L_0} dx ( -  \frac{k^2}{4} y_0(x) \mathcal{O}_{\kappa}^{-1} y_0(x)  + \frac{k  y_0(x)^2}{2}) }.
\end{equation}
The prefactor in the above equation is the partition function of a modified harmonic oscillator (MHO), previously discussed in
Ref.~\cite{Kleinert-path-integrals}. Leaving the details of that aside for the moment, we write free energy of the system as
\begin{equation}
\label{semiflexible-free-energy-preaveraged}
F  = F_{M H O} + T \int_0^{L_0} dx\, \left(- \frac{k^2}{4} y_0(x) \mathcal{O}_{\kappa}^{-1} y_0(x)  + \frac{k  y_0(x)^2}{2}  \right).
\end{equation}

We must now average this free energy over the ensemble of paths taken by the local potential minimum. We average over $y_0(x)$
in an ensemble where we weight each such path by Eq.~\ref{xb-distribution}. Doing so, we obtain
\begin{equation}
\label{semiflexible-free-energy-averaged}
\left[ F \right]  = F_{M H O} -  \frac{T k^2}{8} \mbox{Tr} ( \mathcal{O}_{\kappa}^{-1} G^{-1} )    + \frac{T k  }{4} \mbox{Tr} G^{-1},
\end{equation}
where $G$ is defined by Eq.~\ref{valley-path-weight}.  Once again, we are required to compute the traces of the relevant operators, defined
by Eqs.~\ref{valley-path-weight} and \ref{bending-operator}.  The analogous exercise for the flexible filament
was relegated to appendix~\ref{app:traces}.  We expand on that discussion for the case of semiflexible filaments here.

First we note that one can factor ${\mathcal O}_{\kappa} $ into two commuting operators
\begin{equation}
\label{factorization}
{\mathcal O}_{\kappa} = \frac{\ell_{\rm P}}{2} \left[ \partial^2 - \omega_{1}^{2} \right] \left[ \partial^2 - \omega_{2}^{2} \right] ,
\end{equation}
where we have introduced the (potentially complex) frequencies:
\begin{equation}
\label{frequency-one}
\omega_{1,2}^{2} = \frac{m}{2 \ell_{\rm P}} \left[ 1 \pm \sqrt{1 - \frac{4 \ell_{\rm P} k}{m^{2}} } \right].
\end{equation}
Expanding in a complete eigenbasis of the two operators those product make up ${\mathcal O}_{\kappa}$, we obtain
a form of the disorder averaged free energy in terms of an infinite sum over the (discrete) eigenvalues of ${\mathcal O}_{\kappa}$
indexed by
\begin{equation}
\label{z-n}
z_{n} = n \pi/L_{0},
\end{equation}
where $n =0,1,\dots$.   The resulting free energy is
\begin{widetext}
\begin{equation}
\label{free-energy-stiff-postaveraged}
\left[ F \right]  = F_{M H O} -   \frac{\xi k^2}{2  \ell_{\rm P}} \sum_{n=0}^\infty \frac{1}{  z_{n}^2} \frac{1}{ \left( z_{n}^2 + \omega_1^2 \right) } \frac{1}{ \left(z_{n}^2 + \omega_2^2 \right) }    + \frac{ \xi k  }{2} \sum_{n=0}^\infty \frac{1}{ z_{n}^2  }.
\end{equation}
\end{widetext}
The MHO free energy is given in the appendix~\ref{app:MHO}.
If we take the limit of a vanishing bending modulus ($\ell_{\rm P} \rightarrow 0$), we find that
$\omega_1 \rightarrow \infty $ and  $\omega_2 \rightarrow \sqrt{k/m}$.  This returns us to the previous calculated free energy (up to a constant)
for the  flexible filament in the disordered potential -- see appendix~\ref{app:traces}.  As the bending modulus increases from zero, the two frequencies
become complex when $ \ell_{\rm P} \ge m^{2}/4 k$ -- see Eq.~\ref{frequency-one} .  The free energy, however, remains real since
$\omega_{1,2}^{2}$ are complex conjugates.

Finally, we note that the product of fractions in the above
summation can be broken up into a set of three independently convergent sums.  This allows one to write the disorder-averaged free energy in
terms of a sum of cotangents of $\omega_{1,2}^{2} L_{0}$.  We consider that solution in the
limit of long and stiff filaments $\omega_{1,2} L_0 \gg 1 $, obtaining  a simple algebraic expression
\begin{equation}
\left[ F \right] = F_{\rm M H O} + \frac{1}{2} \frac{k L_0}{2 q}    \frac{ (\omega_2^2  + \omega_1 \omega_2  + \omega_1^2 )  }{ \omega_1 \omega_2 (\omega_1 + \omega_2)   }.
\end{equation}
Using Eq.~\ref{frequency-one}, we reintroduce the original model parameters.  That result is most
succinctly expressed in terms of the dimensionless parameter
\begin{equation}
\label{phi-definition}
\phi =  \sqrt{\ell_{P} \nu}\frac{T}{\xi \tau}.
\end{equation}
In terms of $\phi$ we find the disorder-averaged free energy to be
\begin{equation}
\label{semiflexible-filament-free-energy}
\left[ F \right]  = F_{\rm M H O} + \frac{\sqrt{k m} L_0}{4 q} \frac{ 1 + \phi}{  \sqrt{1 + 2 \phi }} .
\end{equation}

Physically, we see that this dimensionless quantity  $\phi$ is the ratio of the Pincus blob size to the
correlation length (as was observed in the flexible polymer case) multiplied by a correction factor that incorporates the strength of the
random potential.  Specifically, we see that this correction factor is given by the  square-root of the ratio of two length scales in the problem:
the persistence length $\ell_{P}$ of the filament and the arclength of the filament $\nu^{-1} = T/V_{0}$ required for the potential energy of pinning
to equal thermal energy.  We may interpret the effect of a finite  persistence length as extending the size of the underlying Pincus blobs.
This result extends the disorder-averaged free energy of a flexible polymer to the semiflexible regime, and may be compared to the
previous result found in Eq.~\ref{flexible-filament-free-energy}.  The effect of the finite bending modulus enters in both the difference between
$F_{\rm SHO}$ and $F_{\rm MHO}$ in the first term on the right hand side of Eq.~\ref{semiflexible-filament-free-energy} and in second term on the
right hand side of the above equation, where $b$ enters solely through the dimensionless parameter $\phi$ defined in Eq.~\ref{phi-definition}.

We can now calculate the effect of the pinning potential on the length of the chain between the two pinning points.  Following the same approach
and using the same definitions as used for the flexible filament, we now find
\begin{equation}
 \Delta L    =L_{M H O}  + \frac{\xi \sqrt{\frac{k}{m}} L_0}{8 } \frac{1 + 3 \phi }{  \left( 1 + \phi \right)^{3/2}  }
\end{equation}
The pinning potential always increases the arclength, but its contribution becomes smaller as the filament's bending modulus is
increased, {\em i.e.}, for larger $\phi$. Using the result for the MHO and reintroducing the original model parameters, we write
\begin{equation}
\label{semiflexible-filament-length}
\frac{ \Delta L }{L_{0}}  =\frac{T}{4  \sqrt{\tau \kappa}} \left( \frac{1}{ \sqrt{1 + 2 \phi }  } - 1 \right)    + \frac{1}{8}
\sqrt{\frac{V_0}{\tau}} \frac{1 + 3 \phi }{  \left( 1 + \phi \right)^{3/2} }.
\end{equation}
The first term in the above expression is proportional to the ratio of the Pincus blob size $\zeta = T/\tau$ to the {\em bending length}
$\ell_{b} = \sqrt{\kappa/\tau}$, which sets the cross over length between a regime where tension
dominates the statistical ensemble of filament  configurations at longer lengths and the bending modulus at shorter lengths.  The
second term in the above expression is proportional to the ratio of the same Pincus blob size and the length scale set by the pinning
potential $\nu^{-1} = T/V_{0}$.  In the limit of very stiff filaments so that $\ell_{P} \nu \gg 1$ and $\phi \gg 1$ we see that the first term
provides a negative change in length, as observed for the flexible polymer, but the second term increases the filament arclength
with a contribution $\sim \sqrt{\zeta \nu} \phi^{-1/2}$. The interpretation is similar to that given regarding the flexible polymer.  The first
term produces filament straightening as discussed there.  The second term provides for the increase of the arclength due to the tortuosity of
of the potential minima.

\subsection{Semiflexible filaments: force controlled distribution}

We now consider the semiflexible filament to be interacting with the smoother, force controlled distribution.  The calculation proceeds in a manner
analogous to the previous two sections.  The key difference is that the differential operator appearing in the statistical weight of valley paths ${\cal G}$
is replaced by a new one accounting for the fact that the typical curves along bottom of the valleys of the potential now  have their own
persistence length.  Thus the differential operator ${\cal G}$ is replaced by
\begin{equation}
\label{G-force}
{\cal G}_{\rm F} = \frac{\xi}{2} \partial^{4} - \frac{1}{2 \xi} \partial^{2}.
\end{equation}
Following the methods outlined above, we arrive at an expression for the free energy of the filament in the quenched potential, written in
terms of a sum over the eigenvalues of the differential operators ${\cal O}_{\kappa}$ and ${\cal G}_{\rm F}$.  We find that the disorder-averaged
free energy is given by
\begin{equation}
\left[ F  \right] = F_{M H O} + \Delta F,
\end{equation}
where the first term is the free energy of the modified harmonic oscillator, as discussed in appendix B. The second term is the
correction due to complex geometry of the valleys of the potential. It is given by
\begin{equation}
\Delta F  =   \frac{ T k}{2 \xi} \sum_{n=0}^\infty \frac{z_{n}^2   + \omega_2^2 +  \omega_1^2  }{ \left( z_{n}^2 +
\Omega^2 \right)  \left(z_{n}^2 + \omega_1^2 \right) \left(z_{n}^2 + \omega_2^2 \right) },
\end{equation}
where  $\Omega= 1/ \xi$.  The frequencies $\omega_{1,2}$ are defined in Eq.~\ref{frequency-one}
and $z_{n}$ is defined in Eq.~\ref{z-n}.  This result simplifies considerably when we examine the limit of very short or
very  long filaments.  If the former case, where $\omega_{1,2} L_0 \ll 1$, only zeroth term survives and we get
\begin{equation}
\Delta F  =   \frac{T k}{2 \xi} \frac{ \omega_2^2 +  \omega_1^2  }{  \Omega^2  \omega_1^2  \omega_2^2 } .
\end{equation}
In the latter case, where $\omega_{1,2} L_0 \gg 1$, we may change the summation to an integration over $z =  \frac{\pi n}{L_0 }$, obtaining
\begin{equation}
\label{delta-F-integral}
\Delta F  =   \frac{T k L_0 }{2 \xi \pi } \int_{0}^\infty dz \frac{z^2   + \omega_2^2 +  \omega_1^2  }{ \left( z^2 + \Omega^2 \right)  \left( z^2 + \omega_1^2 \right) \left( z^2 + \omega_2^2 \right) }.
\end{equation}
The summation for the general case can also be performed, producing a quite lengthy expression that do not reproduce here.
Performing the integral in  Eq.~\ref{delta-F-integral}, we find
\begin{equation}
\Delta F  =   \frac{T k L_0 }{4 \xi } \frac{ (\omega_1^2 +  \omega_2^2)( \omega_1  + \omega_2)  +  \Omega (\omega_1^2 + \omega_2^2 + \omega_1 \omega_2 )  }{ \Omega \omega_1 \omega_2 (\omega_1 + \omega_2) (\omega_1 + \Omega ) (\omega_2 + \Omega ) }.
\end{equation}

We now wish to compute the average arclength of the filaments to observe the effect of the pinning potential upon their ensemble of
conformations.  This calculation involves taking the derivative of the free energy with respect to the tension, which is conjugate to the
length. First, we write the free energy difference in terms of two auxiliary functions $f$ and $g$ whose argument is $r = m/\ell_{\rm P}$:
\begin{equation}
\Delta F  =  \frac{T \sqrt{k \ell_{\rm P}} L_{0}}{4 }   \frac{f(r)}{g(r)},
\end{equation}
where
\begin{eqnarray}
f(r) &=& \Omega (r + s) + r \sqrt{r + 2 s}\\
g(r) &=&  \sqrt{r + 2 s} (\Omega^2 + \Omega \sqrt{r + 2s} + s )
\end{eqnarray}
and $ s = \sqrt{\frac{k}{\ell_{\rm P}}}$.   Then taking the appropriate derivative, we compute the difference in excess arclength between the filament in
the confining potential and the same filament without it
\begin{equation}
\Delta L = \Delta L_{M H O} + \frac{L_0 }{4} \sqrt{\frac{k}{\ell_{\rm P}} } \frac{f'(r) g(r) - g'(r) f(r)}{g^2(r)} .
\end{equation}
The first term, which is the excess arclength of the semiflexible filament in a uniform harmonic potential, is calculated in appendix~\ref{app:MHO}.

\begin{figure}
\includegraphics[width=8cm]{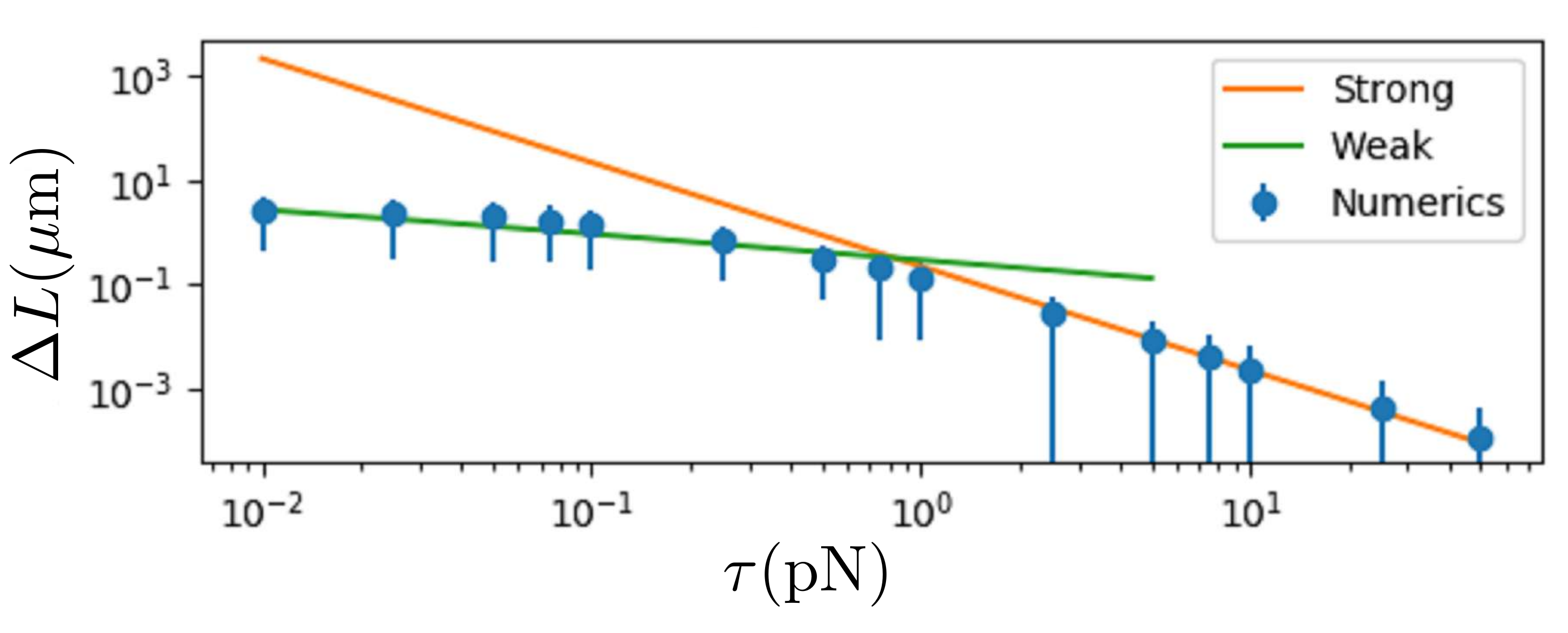}
\caption{(color online) The excess arclength $\Delta L$ -- see Eq.~\ref{definition-excess-length} of a semiflexible filament in the quenched
pinning potential (with persistence length $\ell_{P} \approx 14 \mu m$) as a function of tension $\tau$.  At high tension (Strong, orange) the filament
cannot track the bottom of potential valleys, while at  low tension (Weak, green) or small bending modulus the filament does track the potential valleys
with higher fidelity.  The (filled blue) circles represent
simulation results and the errors show the standard deviations of about five hundred filaments. The pinning
potential is defined by $V_{0} = 0.175 pN$, $\xi \approx 1.6 \mu m$.}
\label{fig:arclength}
\end{figure}

We plot the  $\Delta L$ versus applied tension in Fig.~\ref{fig:arclength}.  There we see the decrease in excess arclength of the filament
with increasing tension $\tau$.  There are two regimes  characterized by a different power laws $\Delta L \sim \tau^{a} + \mbox{const}$ (there is a
finite constant in the low tension regime) in the  low and high tension regimes, referred to as {\em weak} and {\em strong} in the figure caption.
In the weak tension regime, the filament is
better able to track the valley of the potential minimum. As the applied tension is increased, the ensemble of filament configurations
becomes restricted to straighter ones that cannot follow these valleys with high fidelity.

In comparing the theoretical calculation to the numerics, we freely adjusted the curvature and correlation length of the valley to obtain the fit.

\section{Prestress and excess free energy of the pinned filament}
\label{sec:excess}
From the results for the excess free energy, we have a prediction for the increase of the energy density of a network due to the effect of
pinning.  This enhancement of the energy density should be interpreted as the observed {\em pre-stress} found in biopolymer
networks.  Pre-stress can be considered to be stored in at least two separate manners.  First, there should be excess bending and
stretching energy  of the filaments as they are pinned by the network (represented by the quenched potential in our analysis) into
configurations that store  more than $T/2 $ per bending mode -- the amount of energy storage consistent with the equipartition theorem.
Secondly, there may  be excess energy stored in the strain associated with the cross links themselves.  We cannot directly
measure this quantity in our model.  To obtain the excess energy stored in the filament due to the pinning potential, we simply
compute the disorder  average of the thermal expectation value of the squared amplitude of
each Fourier mode of the filament.

\subsection{Strong tension}
First we work under the assumption of strong tension in which the
filament is nearly straight. We start by expanding the undulations of the filament
\begin{equation}
y(x) = \sum_{n=1}^{\infty} u_{n }\sin \left( z_{n} x \right).
\end{equation}
In terms of these Fourier modes the energy in a particular realization of the force-controlled pinning ensemble is
\begin{equation}
\label{Energy-Fourier-modes}
E = \frac{L_{0}}{4} \sum_{n=1}^{\infty} \left\{  \left(\kappa z_{n}^{4}  + \tau z_{n}^{2} \right) u_{n}^{2}+2  f_{n} u_{n}  \right\},
\end{equation}
where $f_{n}$ are the Fourier modes of the pinning force so that
\begin{equation}
\label{Fourier-for-the-force}
f(x) = \sum_{n=1}^{\infty} f_{n }\sin \left( z_{n} x \right).
\end{equation}

In this Fourier expansion of the pinning force, we assume that the
potential is defined in a box of size $[L_x \times L_y ] $ with periodic boundary conditions and we set $L_x = L_0$.
Now, we may use the equipartition theorem to demand that each Fourier mode stores $T/2$ energy.  This leads to
\begin{equation}
\label{u-n-thermal}
\langle u_{n}^{2} \rangle = \frac{ 2 T }{ L_{0 }\left( \kappa z_{n}^{4}+ \tau z_{n}^{2 }\right)} + \frac{ {f_n}^{2}}{\left( \kappa z_{n}^{4}+ \tau z_{n}^{2 }\right)^{2}}.
\end{equation}

We now average the above result using the force-controlled distribution for the potential from Eq.~\ref{Force-Distribution} in Fourier representation
\begin{widetext}
\begin{equation}
\label{Distrib-for-pot-Fourier}
P(V_{k_x,k_y}) \propto \exp \left\{ -L_x L_y  V_{k_x,k_y}^2  \left( \frac{\xi^2}{8 V_{0}^2}  (k_x^2 + k_y^2)^2   + \frac{1}{8 V_{0}^2} (k_x^2 + k_y^2) \right) \right\}.
\end{equation}
\end{widetext}
where Fourier modes are defined in a standard way
\begin{equation}
V(x,y)=\sum_{k_x,k_y} V_{k_x,k_y} \sin(k_x x) \sin(k_y y).
\end{equation}
We now express $f_n$ from Eq.~\ref{Fourier-for-the-force} in terms of the gradients of $V_{k_x, k_y} $
\begin{equation}
\label{Force-from-the-pot}
f_n = \sum_{k_y} k_y V_{k_x,k_y}
\end{equation}
with $k_x = \frac{\pi n}{L_0} = z_{n}$
Using Eq.~\ref{Force-from-the-pot} and Eq.~\ref{Distrib-for-pot-Fourier} we obtain
\begin{equation}
[f_n^2] = \sum_{k_y} \frac{k_y^2}{2 L_x L_y  \left(  \frac{\xi^2}{8 V_{0}^2}  (k_x^2 + k_y^2)^2   + \frac{1}{8 V_{0}^2} (k_x^2 + k_y^2) \right) }.
\end{equation}
Taking $L_y$ to be large we replace the summation with integration to obtain
\begin{equation}
\label{force-variance}
[f_n^2] = \frac{2 V_0^2}{ L_x \xi^2 \left( \sqrt{z_n^2 + \frac{1}{ \xi^2} } + z_n \right) }.
\end{equation}

Since only the second term in Eq.~\ref{u-n-thermal} depends on the pinning potential the average over the quenched disorder yields
an expression for the mean excess bending energy of the filament
\begin{equation}
\label{bending-answer}
E^{\rm bend}_n =  \frac{  T \kappa z_{n}^4   }{ 2(\kappa z_{n}^4  + \tau  z_{n}^2)} + \Delta E^{\rm bend}_n .
\end{equation}
in which the second term contains all the information about the filament's interaction with the pinning potential.  That second term is
\begin{equation}
\label{mode-storage}
\Delta E_n = \frac{  \kappa z_{n}^4  V_{0}^2 }{ 2 (\kappa z_{n}^4  + \tau  z_{n}^2)^2  \xi^2 \left( \sqrt{z_n^2 + \frac{1}{ \xi^2} } + z_n \right)} .
\end{equation}
The first term is merely the standard result from the equipartition theorem for a semiflexible filament~\cite{MacKintosh:95}.  It follows
similarly that contribution from the potential to the mean energies stored in filament tension and in the pinning potential are given by
\begin{eqnarray}
\label{tension-energy-storage}
\Delta E^{\rm ten}_n &=& \frac{1}{\ell_{t}^{2} z_{n}^{2}} \Delta E^{\rm bend}_n\\
\label{pot-energy-storage}
E^{\rm pot}_n &=& \frac{k}{\kappa z_{n}^{4}} \Delta E^{\rm bend}_n.
\end{eqnarray}

\begin{figure}
\includegraphics[width=8cm]{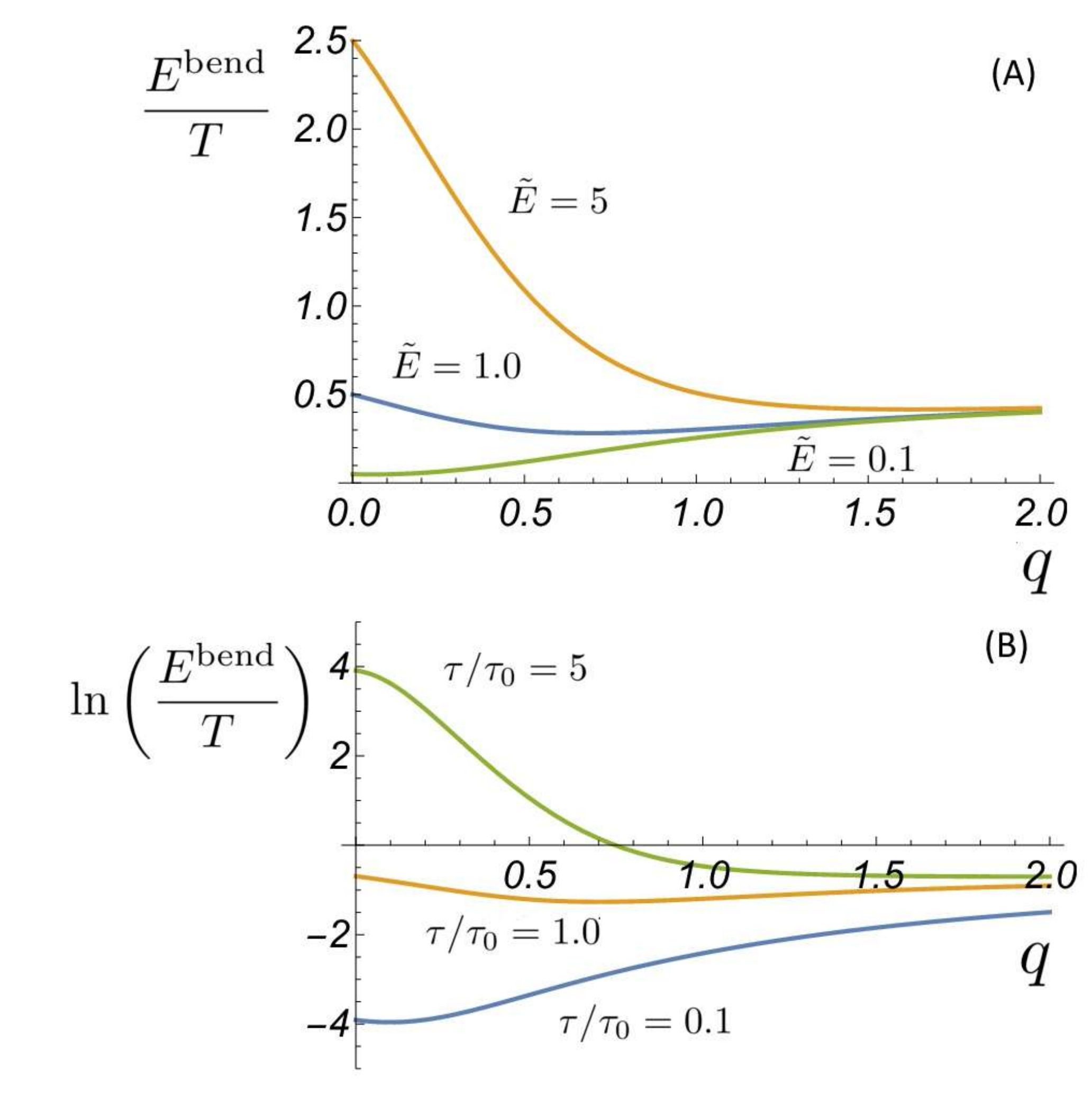}
\caption{(color online) (A) $E^{\rm bend}(q)$, the energy energy per mode of a tensed, pinned filament
(in units of $T$) as a function of $q = \xi z_{n}$.  The
pinning potential strength is set by dimensionless $\tilde{E} = V_{0}^{2} \xi/T \tau$ using the exponential potential distribution. The small $q$
modes typically have more bending energy than expected for a thermalized filament without the pinning.  The high $q$ modes are effectively
unpinned. (B) The effect of changing tension on bending energy: $\tau = 0.1,1,5 \tau_{0}$ where $\tau_{0} V_{0}^{2} \xi/T$. We set $\kappa = \xi^{2}
\tau_{0}$.}
\label{fig:pinning}
\end{figure}

In Fig.~\ref{fig:pinning} we plot the bending energy stored in the filament as a function of dimensionless wavenumber $q = \xi z_{n}$ for a
variety of pinning potential strengths (at fixed external tension) (upper panel A) and a variety of tensions at a fixed value of the strength of the
pinning potential (lower panel B).  From dimensional analysis we note that there is a single scale that sets the strength of the pinning potential
\begin{equation}
\tilde{E} = \frac{V_{0}^{2} \xi}{T \tau}.
\end{equation}
When this quantity is large $\tilde{E} \gg 1$ we expect the pinning potential to control  the statistical ensemble of the filament configurations.
Conversely, we expect high tension to straighten out the filament so that it cannot
follow the local potential minima.  Higher tension leads to both straighter typical filament configurations and configurations for which the
effect of pinning becomes harder to distinguish against a background of thermal undulations.
This transition between pinning dominated states of the filament and thermally dominated ones is
wavenumber dependent.
At sufficiently high wavenumbers $q = \xi z_{n} > q^{\star}$ the modes of the filament are generically freed from the pinning potential.

\begin{figure}[h]
\includegraphics[width=8cm]{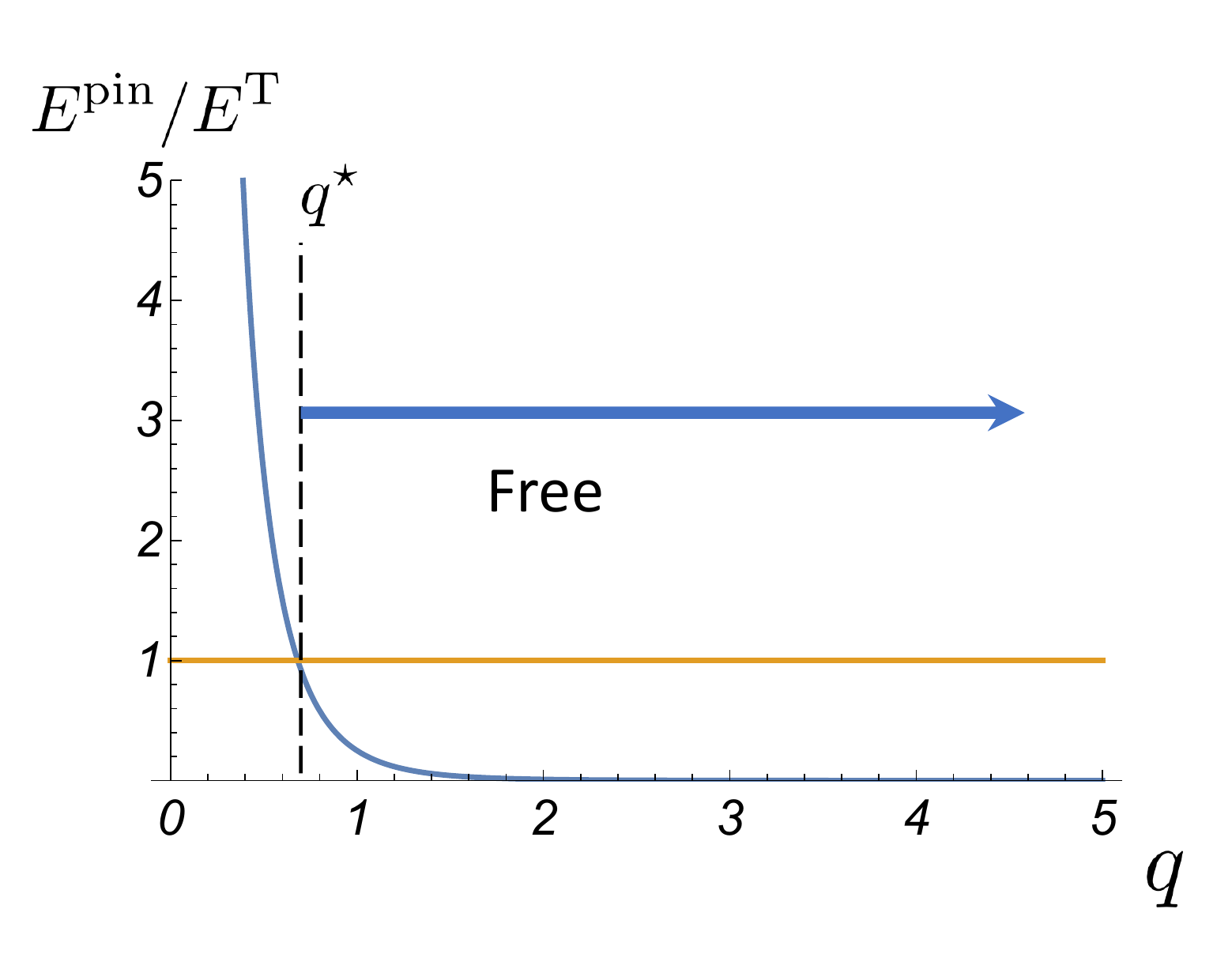}
\caption{(color online) For a given value of the pinning strength and tension, there is a transition at $q^{\star}$
between pinned modes $q < q^{\star}$,  which trap a significant excess energy as compared to the free filament and free
modes $q < q^{\star}$, which do not.  We examine this transition by plotting the ratio of the excess bending energy resulting from
the pinning potential $E^{\rm pin}$ to the energy of that mode without the pinning potential $E^{\rm T}$.  The pinning potential and the tension
are $T/(2 \xi)$ and $ \sqrt{2 \kappa V_0}/\xi$.  The figure is qualitatively the same for other values of these parameters. }
\label{fig:freedom}
\end{figure}

In Fig.~\ref{fig:freedom} we observe this transition from strongly pinned modes, where the pinning contribution to the bending energy
$E^{\rm pin}$ is greater than the thermal component $E^{\rm T}$ for $q <  q^{\star}$ to effectively unpinned ones.

Since the work done by tension to extend the filament is $\tau \Delta L$, we can use  Eqs.~\ref{mode-storage},\ref{tension-energy-storage}
to compute the excess length
\begin{equation}
\Delta L = \sum_{n=1}^\infty  \frac{    V_{0}^2 }{  2 (\kappa z_{n}^2  + \tau )^2 z_n^2  \xi^2 \left( \sqrt{z_n^2 + \frac{1}{ \xi^2} } + z_n \right)}.
\end{equation}
The sum is rapidly convergent so the first terms will dominate. In the limit of a short correlation length
$\xi/L_0 \ll 1$, $ L_0 \sqrt{\frac{\tau}{\kappa}} \gg 1$ we get
\begin{equation}
\label{Delta-L-large-tension}
\Delta L = \sum_{n=1}^\infty  \frac{   V_{0}^2 }{  2 \tau^2  z_{n}^2  \xi }.
\end{equation}
The summation results in
\begin{equation}
\Delta L = \frac{   V_{0}^2 L_0^2 }{ 12 \tau^2   \xi }.
\end{equation}

 If instead we consider the case of the large bending we should take the limit $L_0 \sqrt{\frac{\tau}{\kappa}} \ll 1 $ and

 \begin{equation}
 \label{Delta-L-large-bending}
 \Delta L = \sum_{n=1}^\infty \frac{V_0^2}{2 \kappa^2 z_n^6 \xi} = \frac{V_0^2 L^6}{1890 \kappa^2 \xi}
\end{equation}

For the case of large correlation length, $\xi/L_0 \gg 1$, large tension $L_0 \sqrt{\frac{\tau}{\kappa}} \gg 1 $ we get
\begin{equation}
\Delta L = \sum_{n=1}^\infty  \frac{   V_{0}^2 }{ 2 \tau^2  z_{n}^3   \xi^2    },
\end{equation}
which results in
\begin{equation}
\Delta L =  \frac{   V_{0}^2 L_0^3 }{ 4 \pi^3\tau^2    \xi^2    }   \zeta(3)
\end{equation}
where $\zeta(x)$ is Riemann $\zeta$-function. Since the
correlation length is much larger than that the filament, we expect that the filament feels an essentially uniform force field, much like that of a
hanging rope in a gravitational field.  Indeed, the result for the excess length in this classical problem is $\Delta L =
\frac{g^2 L_0^3}{6 \tau^2} $, which demonstrates the same power law dependence on tension $\tau$ and the separation of the end points $L_{0}$.

\subsection{Weak tension}

For case of weak tension where the filament can better follow the potential minima, we are free to use the valley approximation.
In this case the filament's energy is given by Eq.~\ref{semiflexible-energy}.  Doing the same calculation for the variance of each
Fourier mode of the filament in a valley whose bottom curve is described by the Fourier modes of $y_0(x)$, $y_0^{n}$, we find
\begin{equation}
\label{u-n-thermal-weak-tension}
\langle u_{n}^2 \rangle =   \frac{2 T}{L_{0} \left(\kappa z_{n}^4  + \tau  z_{n}^2 + \frac{V_{0}}{ \xi^2} \right)} +
  \frac{V_{0}^2 (y_0^{n})^2}{ \xi^{4}\left(\kappa z_{n}^4  + \tau  z_{n}^2 + \frac{V_{0}}{ \xi^2} \right)^{2}}.
\end{equation}

Once again, the first term is independent of the disorder in the valley, but does depend on the curvature of the potential.   This result
corresponds to the case of a semiflexible filament in a straight parabolic potential~\cite{Smith:01}.  The second
term corrects this result for the tortuosity of the valley.  To compute this correction we note that
\begin{equation}
\label{disorder-average-variance-y-b}
\left[ (y_0^n)^2 \right] = \frac{1}{\frac{L}{2 \xi} z_{n}^2 + \frac{L \xi}{2} z_{n}^4}.
\end{equation}
From this result and from energy function -- see Eq.~\ref{semiflexible-energy}, we immediately find that
\begin{widetext}
\begin{equation}
\left[ E^{\rm bend}_n \right ]  =  \frac{    T  \kappa z_{n}^{4}}{ 2 (\kappa z_{n}^4  + \tau  z_{n}^2 + \frac{V_{0}}{2 \xi^2}) } +
\frac{\kappa V_{0}^2 z_{n}^{4}}{ 4 \xi^4 \left(\kappa z_{n}^4  + \tau  z_{n}^2 + \frac{V_0}{ \xi^2}\right)^2   \left( \frac{1}{2 \xi}z_{n}^2 + \frac{\xi}{2} z_{n}^4 \right)}.
\end{equation}
\end{widetext}

\begin{figure}[h]
\includegraphics[width=8cm]{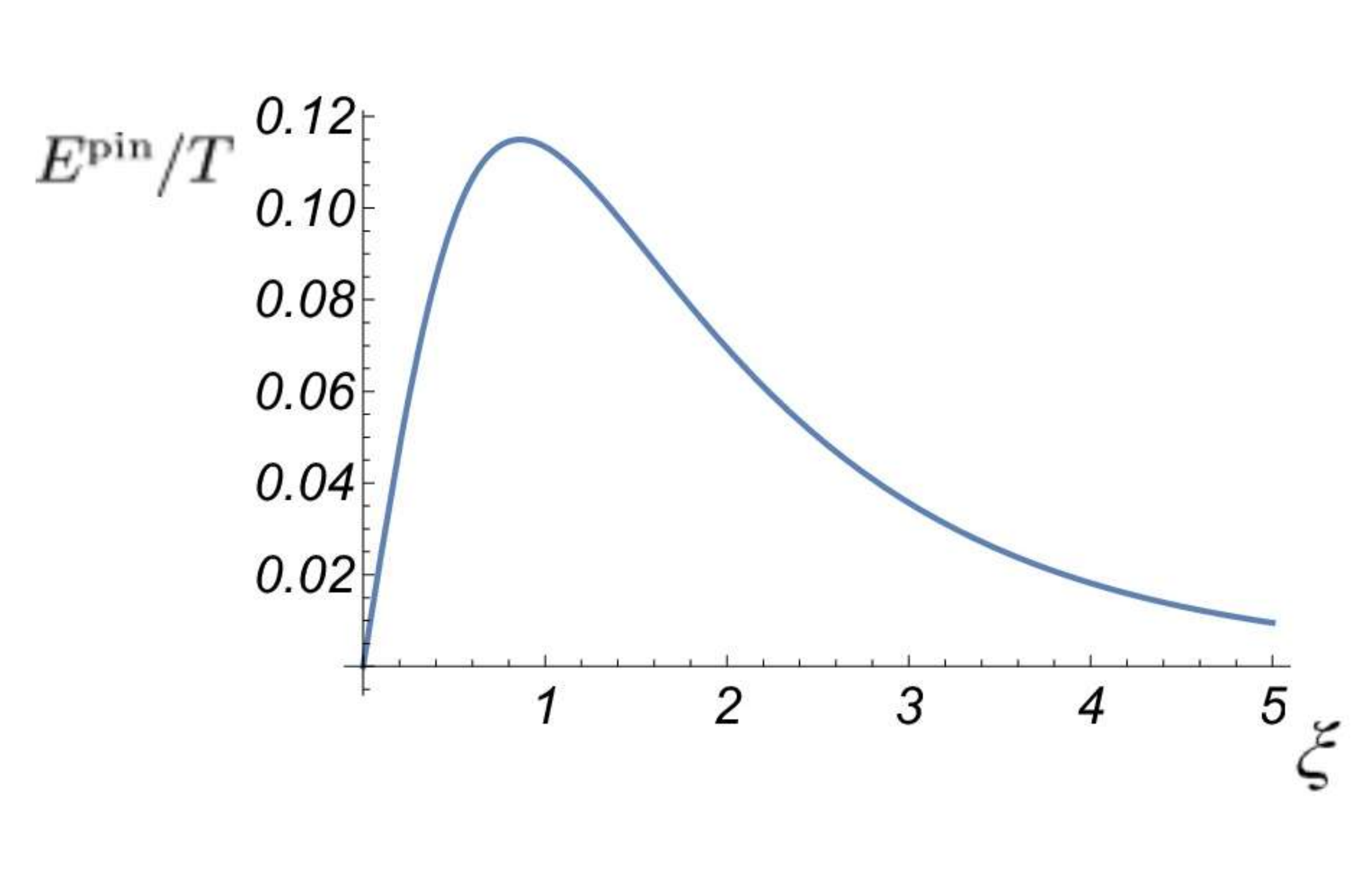}
\caption{The maximum value for the energy stored in the potential is reached at $\xi = 1/z_n$ which resembles the resonance absorption spectrum. Here $\xi$ is measured in the units of $1/z_n$, $V_0 = 20(\tau + \kappa) $ in these units.   }
\label{fig:E-xi}
\end{figure}

Again the first term represents the bending energy associated with the semiflexible filament in a straight parabolic potential. The potential
decreases the bending energy, as is physically reasonable. The potential suppresses the
normal thermal undulations of the filament. The second term, however, represents an increase in the mean bending energy associated with
the curvature of the potential valleys. For a fixed local mean curvature of the pinning potential, $V_{0}/\xi^{2} = \mbox{const}$, the dependence
of this tortuosity correction is nonmonotonic in wavenumber.  We examine this in Fig.~\ref{fig:E-xi}.  There we see that the wavenumber
dependent bending energy goes through a local maximum at the scale of the potential's correlation length,{\em i.e.}, where $z_{n} \xi = 1$.

\section{Simulations}
\label{sec:simulations}
The established computational framework for the Brownian dynamics of semiflexible filaments proposed in~\cite{Cyron2012} and used e.\,g.~in \cite{Levine:14} has been extended to account for forces resulting from the random potential field.

\subsection{The finite element Brownian dynamics simulation framework}
A single filament is modeled by nonlinear, geometrically exact, 3D Simo-Reissner beam theory and discretized in space using finite elements~\cite{Jelenic1999}.
In terms of the structural rigidity of the filament, we thus account for axial, torsional, bending, and shear deformation.
To model the Brownian motion, we include viscous drag as well as thermal forces, each distributed along the entire filament length.
More precisely, viscous forces and moments are computed assuming a quiescent background fluid and individual damping coefficients for translations parallel and perpendicular to the filament axis as well as rotation around the filament axis, respectively.
Thermal forces are determined from the stochastic Wiener process in accordance with the fluctuation-dissipation theorem.
Finally, an Implicit-Euler scheme is used to discretize in time and a Newton-Raphson algorithm solves the resulting nonlinear system of equations.
Further details on this simulation framework including all formulae can be found in~\cite{Cyron2012}.

\subsection{Incorporation of the background potential field}
As described in section~\ref{sec:Introduction}, the potential field~$V$ acting on the filament has dimensions of energy per length.
Its contribution to the virtual work required for the weak, variational formulation of the problem can be stated as
\begin{equation}\label{eq:sim_virtual_work_pot}
  \delta \Pi = - \int_0^{L_0} ds \left\{ \left( \nabla V(\mathbf{r}(s)) \right)^T \, \delta \mathbf{r}(s)  \right\}
\end{equation}
where~$\mathbf{r} \in \mathbb{R}^3$ is the centerline position and~$s \in [0, L_0]$ denotes the arclength coordinate in the stress-free reference configuration of the filament.
Subsequent discretization of the admissible centerline variations~$\delta \mathbf r$ according to the finite element method yields the contributions to the discrete element force vector.
We apply the trapezoidal rule on each finite element to numerically evaluate the integral along the filament.
Regardless of the fact, that we only consider planar problems throughout this article, the entire simulation framework as well as Eq.~\ref{eq:sim_virtual_work_pot} is capable of modeling arbitrary filament configurations in 3D.
Note that the potential exerts forces on the filament, however, as it models a surrounding network, it is independent of the filament motion.
This is commonly denoted as one-way coupling.

To mention the most important algorithmic details:
In a pre-processing step, we use a random number generator, apply a discrete Fourier transformation, and finally a finite difference scheme to arrive at the force field~$-\nabla V$ on a sufficiently fine grid in the entire simulation domain.
In each iteration, we then interpolate these grid values to compute~$-\nabla V$ at the current position of each node and evaluate Eq.~\ref{eq:sim_virtual_work_pot} element-wise.

\subsection{Simulation setup and results}
%
\begin{figure}[htb]
  \subfigure[]
  {
  \includegraphics[width=8cm]{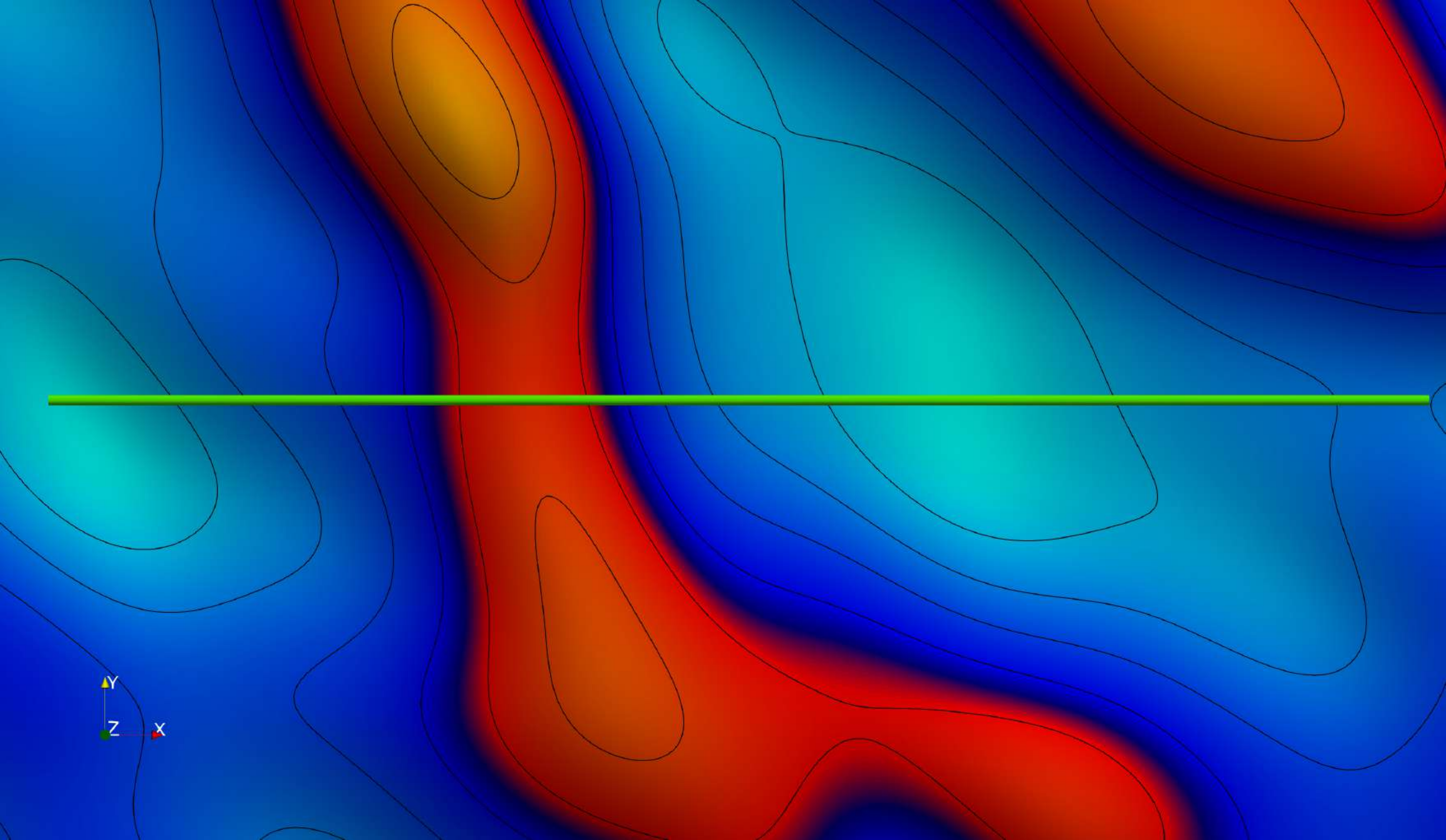}
  \label{fig:sim_setup}
  }
  \subfigure[]{
  \includegraphics[width=8cm]{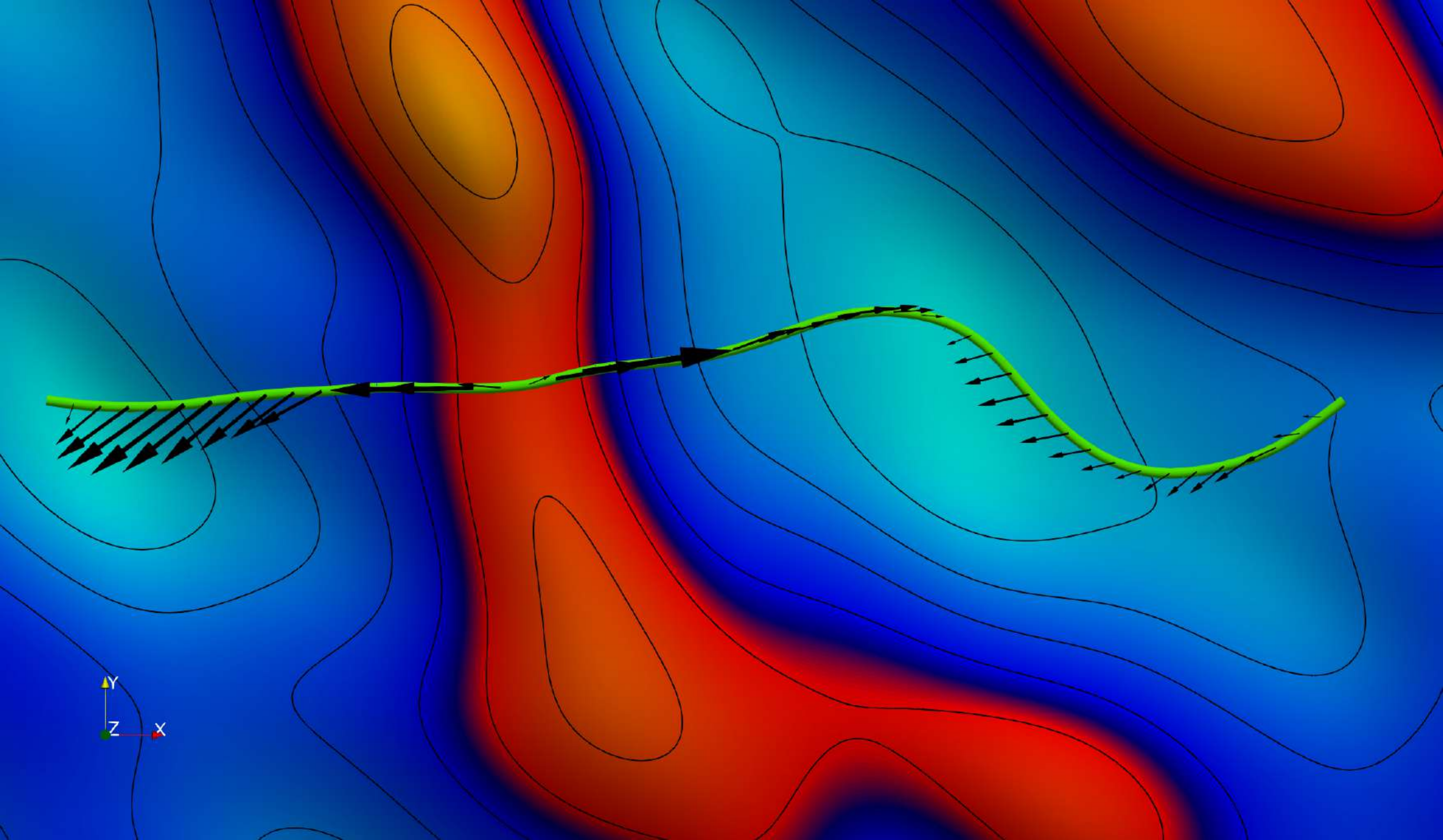}
  \label{fig:sim_snapshot}
  }
  \caption{(color online) (a) Simulation snapshot of the initial setup. An initially straight, stress-free filament is constrained to the $xy$-plane and simply supported at its endpoints. It interacts with a random potential~$V(x,y)$ that is shown as a heat map with contour lines. (b) Simulation snapshot of a deformed configuration showing the forces on the filament resulting from the pinning potential.}
\end{figure}
%
The simulation setup consists of a single filament of length $L_0=20 \mathrm{\mu m}$ and persistence length~$L_p \approx 14 \mathrm{\mu m}$.
Its initial, stress-free reference configuration is chosen straight and parallel to the global $x$-axis, as shown in Fig.~\ref{fig:sim_setup}.
By means of Dirichlet boundary conditions, the filament is constrained to the~$xy$-plane and simply supported, i.\,e., free to rotate at both ends, however only free to move in $x$-direction at one endpoint.
Its circular cross-section is specified by the area~$A=1.9 \times 10^{-5}\mathrm{\mu m}^2$, area moment of inertia~$I = 4.3{\times}10^{-12}\mathrm{\mu m}^4$ and polar moment of inertia~$I_p = 8.6{\times}10^{-12}\mathrm{\mu m}^4$.
The material is defined by the Young's modulus~$E=1.3\times10^{10} \mathrm{pN/\mu m^2}$ and the Poisson ratio~$\nu=0.3$.
Temperature is set to $T{\,=\,}293\mathrm{K}$ and the dynamic viscosity of the quiescent background fluid to $\eta{\,=\,}10^{-3}\,\mathrm{Pa\,s}$.
The filament is discretized in space using $400$ linear beam finite elements and the time step size is chosen as~$\Delta t = 0.01 \mathrm{s}$.

Two variants of the potential field have been considered in simulations.
First, the potential with exponential suppression of high Fourier modes defined by Eq.~\ref{Exponential-suppression} and second, the force controlled distribution defined by Eq.~\ref{Force-Distribution}.
\subsubsection{Results for the potential with exponential suppression of high wavenumber modes}
For the results presented already along with the theoretical prediction in Fig.~\ref{fig:arclength}, we applied the potential from Eq.~\ref{Exponential-suppression}, using $3096$ Fourier modes, a correlation length of~$\xi \approx 1.6 \mathrm{\mu m}$ and $V_0 = 0.175 \mathrm{pN}$.
In addition, a single point force~$\tau = 10^{-2} \ldots 50 \mathrm{pN}$ was applied in global $x$-direction to the (right) endpoint of the filament that is free to move in this direction.
Each simulation was run for $5 \times 10^4$ time steps.
To speed up simulations, we made use of parallelization and simulated systems of five filaments at a vertical spacing much larger than~$\xi$ and without any interactions between the filaments.
Each data point in Fig.~\ref{fig:arclength} results from the statistical ensemble of~$70$ to $100$ such systems with five filaments each, depending on the deviation in results that was higher for the small tension values.
Finally, the excess arclength~$\Delta L$ is obtained from simulation data as the negative displacement of the (right) filament endpoint in $x$-direction.
\subsubsection{Results for the force controlled distribution}
While the simulations using the potential with the exponential suppression of the high wavenumber modes are more robust, the theory assumes a force controlled
distribution.
Therefore, unlike for the previous variant of the potential distribution, there are no adjustable parameters necessary to directly compare these numerical results to the theory.
We consider two parameter sets: one representing the case of large tension and another one representing the case of large bending and zero tension.

The first parameter set is given as $V_0 = 1 / 256^2$ pN, $\tau = 0.006$ pN, $\xi = 1 \mu$m, $L_0 = 5 \mu$m.
The simulation gives us $\langle \Delta L \rangle = 2.0 \pm 0.6 \times 10^{-5} \mu$m, while Eq.~\ref{Delta-L-large-tension} predicts $\Delta L \approx 1.3 \times 10^{-5} \mu $m.
The second parameter set is $V_0 = 1 / 256^2$ pN, $\tau = 0$, $\xi = 1 \mu$m, $L_0 = 5 \mu$m, $\kappa \approx 0.0125 \, \mbox{pN} \, \mu \mbox{m}^2 $.
Here, the result from our simulations is $\langle \Delta L \rangle = 2.2 \pm 0.4 \times 10^{-5} \mu$m, while Eq.~\ref{Delta-L-large-bending} predicts $\Delta L \approx 1.2 \times 10^{-5} \mu $m.
We performed sixteen simulation runs for each parameter set.

\section{Discussion}
\label{sec:discussion}
We have examined the statistical mechanics of a single semiflexible filament in a quenched pinning potential as a model for studying how
the network environment changes the typical stored elastic energy of filaments and leads to prestress.  Based on these calculations we
propose that there are two experimental quantities for which we may make predictions even with our single filament model.  The first is that we
expect the pinning environment of the network to impose a different (and nonequilibrium) statistical weight to filament configurations. One way
to parameterize this difference between the ensemble of filament configurations in a network and of a filament in isolation is that the effective
persistence length of the network filament will no longer be $\ell_{\rm P} = \kappa/T$.

\begin{figure}[h]
\includegraphics[width=8cm]{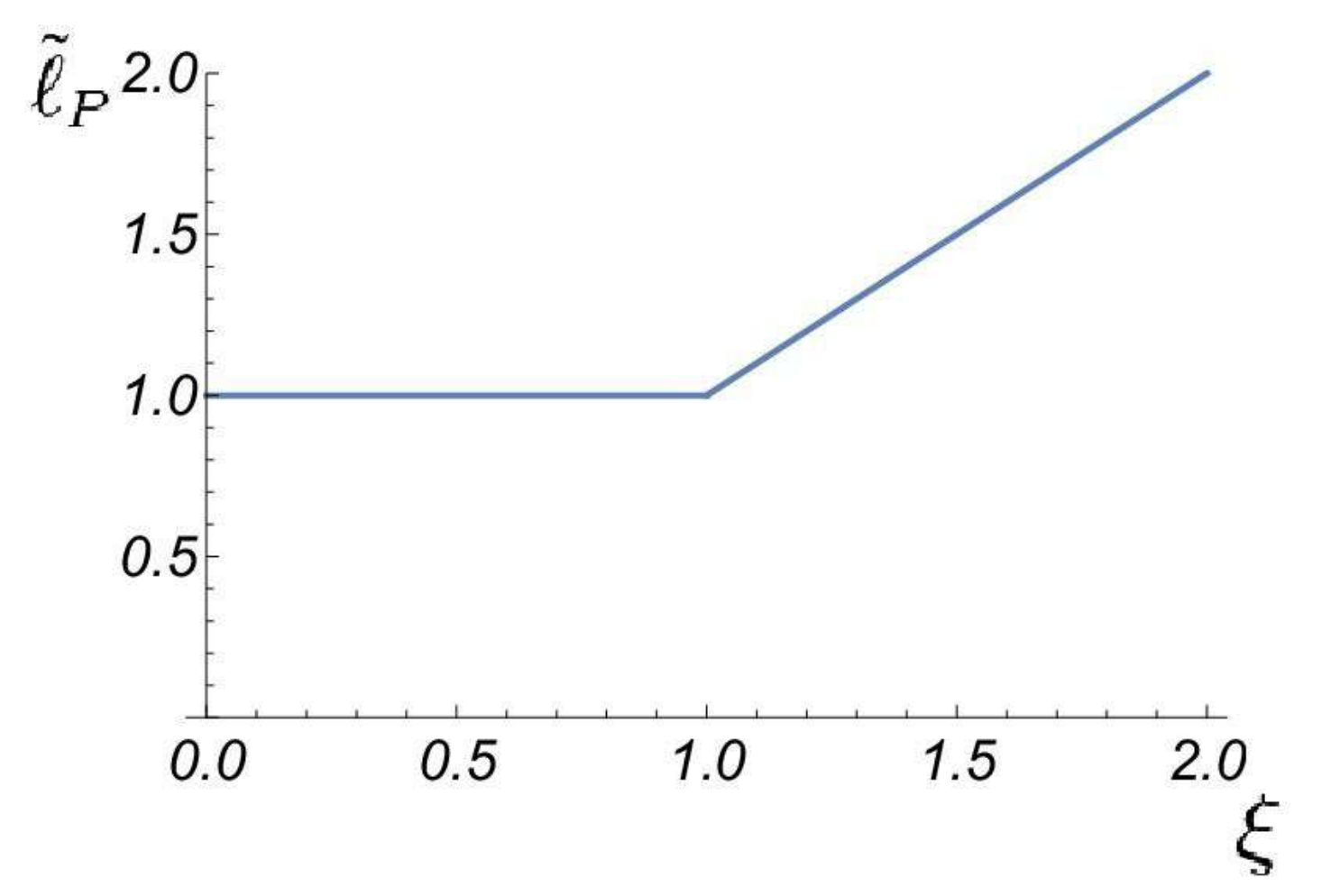}
\caption{Weak tension persistence length}
\label{fig:Weak-tension-persistence}
\end{figure}

Using our results for the disordered-averaged Fourier modes of the filament's undulations, we may directly compute the
tangent tangent correlations. We find that
\begin{equation}
\label{nonthermal-persistence-length}
G(x_1,x_2) = \langle \dot{y}(x_1) \dot{y}(x_2) \rangle  \propto e^{- |x_1 - x_2 |/\tilde{\ell}_{\rm P} },
\end{equation}
where the {\em nonthermal} persistence length $\tilde{\ell}_{\rm P}$ is given by
\begin{equation}
\label{nonthermal-persistence-length-2}
\tilde{\ell}_p = \mbox{max}\left\{ \frac{1}{\mbox{Re}(\omega_{1})}, \frac{1}{\mbox{Re}(\omega_{2})} , \xi \right\},
\end{equation}
where $\omega_{1,2}$ are the eigenvalues introduced in Eq.~\ref{frequency-one}. We examine the dependence of the effective,
disorder-influenced persistence as a function of tension and potential correlation length in Fig.~\ref{fig:Weak-tension-persistence}.

\begin{figure}[h]
\includegraphics[width=8cm]{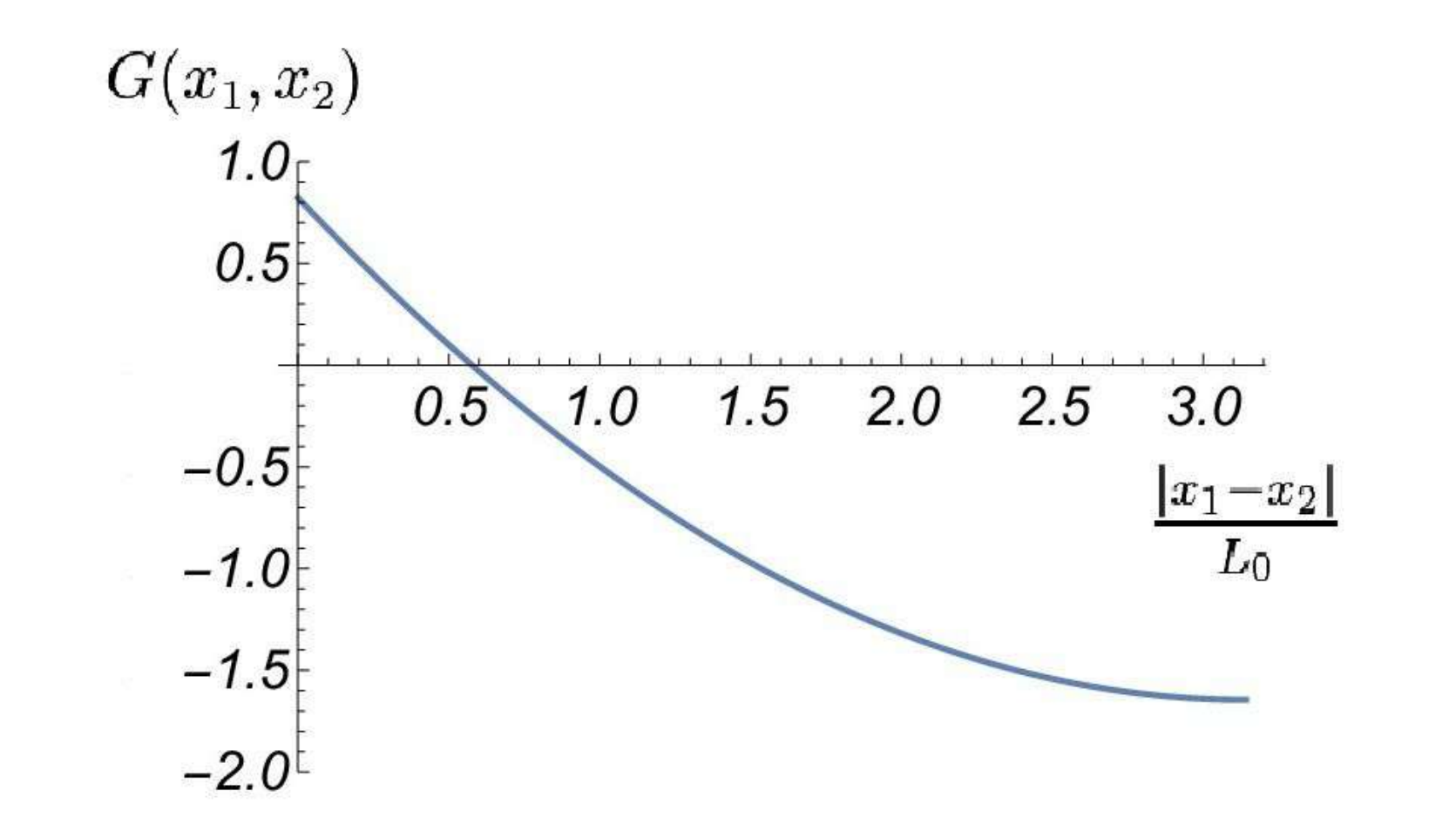}
\caption{Strong tension correlation function}
\label{fig:Strong-tension-cor-fun}
\end{figure}

For the case of sufficiently strong tension or weak pinning potential $\tau \gg V_0$, we find that the
tangent correlations do not decay exponentially with separation along the filament.  Instead  they decay as a polylogarythmic
function of the separation: $\mbox{Li}_2 (e^{i \pi |x_1 - x_2| /L }) $.  In that case, no persistence length can be defined.
See Fig.~\ref{fig:Strong-tension-cor-fun} for the behavior of the correlation function

Turning to the case of prestress, we offer a prediction for the mean excess free energy density of the network due to trapped elastic deformations
of its constituent filaments.  To arrive at this prediction within our one filament model, we assume the excess free energy density may be computed
by summing the excess bending and tension energy $\Delta E^{\rm bend} + \Delta E^{\rm ten}$ of a filament due its interaction with the
pinning potential and then dividing that quantity by the area occupied by that filament.  In
our two dimensional calculation this is simply $L_{0} \times \xi$.  To make a prediction for a three dimensional network we assume that there are
two independent polarization states of the filament's undulations (which is reasonable for small bending angles) resulting in a prediction
\begin{equation}
\label{prestress}
\Delta F \simeq \left( 2 \Delta E^{\rm bend} + \Delta E^{\rm ten}\right) L_{0}^{-1} \xi^{-2}.
\end{equation}
We expect that this quantity should set the scale for the anomalous nonequilibrium stress fluctuations observed in transiently cross linked networks
of semiflexible filaments.

\begin{figure}[h]
\includegraphics[width=8cm]{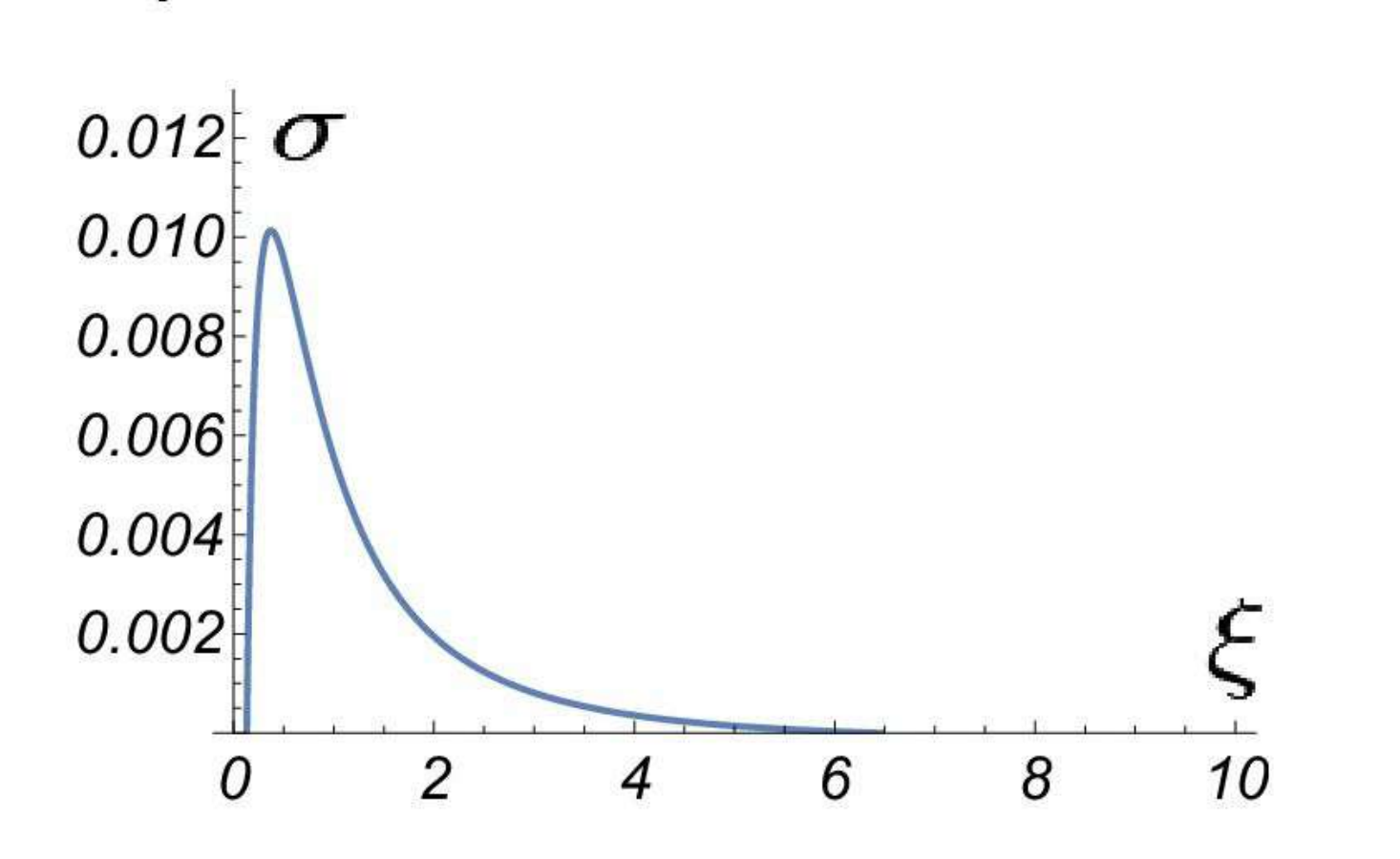}
\caption{Prestress }
\label{fig:Prestress}
\end{figure}

There are limits to the single filament description of the collective phenomenon of network structure and excess free energy.  Following the example
of mean field models of magnets, one might imagine finding a self consistent description of network structure in which the statistics of the pinning
potential are determined by the calculated properties of a filament in that potential in a sort of Weiss molecular field description.  We leave such
self-consistent calculations to future work.

We also note that our description of a single filament is inadequate for studying the crumpling of filaments in a even stronger pinning environments.
States of larger deformation, as might be expected in flexible polymers without sufficient tension, cannot be described by our framework; their
description requires more complex and inherently nonlinear elasticity.  Fortunately, nature provides numerous examples of semiflexible protein
filaments for which our analysis should be sufficient.  Transiently cross linked networks of such stiff filaments are an arena for the study of
the role of quenched disorder on their ensemble of shapes and elastic energy storage.

\acknowledgements
AJL and VMS acknowledge partial support from NSF-DMR-1709785. AJL and VMS are thankful to Robijn Bruinsma and Jonathan Kernes for fruitful discussions.  
WAW and MJG acknowledge partial support from BaCaTeC.

\appendix

\section{Calculating the traces}
\label{app:traces}
In this appendix we discuss the calculation of the trace of the operator $\mathcal{O}^{-1} \mathcal{G}^{-1}$ obtained
in Eq.~\ref{free-energy-change}.  The operators $\mathcal{O}$ and $\mathcal{G}$ were introduced in
Eqs.~\ref{flexible-operator} and ~\ref{valley-path-weight} respectively.
To calculate the traces, we need the complete spectrum of the two differential operators in question. In other words, we must solve
\begin{equation}
\mathcal{O} \Psi = E_n \Psi
\end{equation}
subject to the fixed-end boundary conditions $\Psi(0) = \Psi(L_0)$. The solution is a constant and  the set of
standing waves satisfying the boundary conditions: $\Psi(y) = A \sin( z_{n} y)$, where, as discussed in the main text, $z_{n}= n \pi/L_{0}$
for nonnegative integer $n$.  Defining $\omega = \sqrt{\frac{k}{m}} $ we may write the trace as
\begin{equation}
 \mbox{Tr} G^{-1}  = 2 \xi  \sum_{n=0}^\infty \frac{1}{z_{n}^2 }.
\end{equation}
Since both operators can be simultaneously diagonalized, it is also possible to write
\begin{equation}
\mbox{Tr} ( \mathcal{O}^{-1} G^{-1} ) =   \frac{4 \xi}{ k } \sum_{n=0}^\infty \frac{1}{  z_{n}^2 } \frac{1}{  z_{n}^2 + \omega^{2}  }
\end{equation}
Using the above results and Eq.~\ref{free-energy-change}, one finds directly that
\begin{equation}
\Delta F = \frac{m \xi}{2 }  \sum_{n=0}^\infty \frac{1}{z_{n}^2 }  \left( 1- \frac{\omega^{2}}{  z_{n}^2 + \omega^{2} }  \right).
\end{equation}
The second term in the product occurring in the summand removes the singularity associated with the $n=0$ term.
\begin{equation}
\Delta F = \frac{k \xi}{2 }  \sum_{n=0}^\infty \frac{1}{ z_{n}^2 + \omega^{2}  }
\end{equation}
The summation can be done in closed form leading to the disorder-averaged free energy:
\begin{equation}
 \left[F \right]  = F_{S H O} +  \frac{k \xi}{4 }  \frac{   1  +  \omega  L_0 \coth(\omega L_0)  }{\omega^2 }
\end{equation}

\section{The Modified Harmonic Oscillator}
\label{app:MHO}
In this appendix we review the partition function of the modified harmonic oscillator.   The modified harmonic
oscillator Hamiltonian has an addition $p^{4}$ where $p \rightarrow -i \partial$ is the momentum operator.  This leads to
a time-independent Schr\"{o}dinger operator of the form $\mathcal{O}_{\kappa}$ introduced to discuss the semiflexible filament in the
text. A more complete exposition of this problem can be found in Refs.~\cite{Kleinert-path-integrals}
and \cite{Feynman-book}.  We begin by factoring the
differential operator $\mathcal{O}_{\kappa}$, defined in Eq.~\ref{bending-operator} into a product of two commuting second order
differential operators as shown in Eq.~\ref{factorization}:
\begin{equation}
\label{factorization-2}
{\mathcal O}_{\kappa} = {\mathcal O}_1  {\mathcal O}_2,
\end{equation}
where
\begin{equation}
{\mathcal O}_{j} = \partial^{2}+ \omega^{2}_{j}.
\end{equation}
It immediately follows that the partition sum for the modified harmonic oscillator is given by a product of the partition functions of
two separate harmonic oscillators $Z_1$ and $Z_2$,
\begin{equation}
\label{Z-MHO}
Z_{M H O }  =\frac{1}{\sqrt{\det \mathcal{O}_1}} \frac{1}{\sqrt{\det \mathcal{O}_2}} = Z_1 Z_2,
\end{equation}
with frequencies $\omega_1$ and $\omega_2$ respectively.  The free energy immediately follows.  From Eq.~\ref{Z-MHO}, it
is the sum of the free energies of the two harmonic oscillators introduced above in Eq.~\ref{factorization-2},
\begin{equation}
\label{MHO-free-energy}
F_{M H O }  = F_1 + F_2 = T \ln (2 \sinh (\omega_1 L_0 /2) ) + T \ln (2 \sinh (\omega_2 L_0 /2) ).
\end{equation}
The remainder of this appendix uses the above result.

We first compute the free energy of the filament in a confining potential (finite $k$).
Since the mass term in the equation of motion represents tension $m = \tau/T$ -- see Eq.~\ref{flexible-operator}-- and since tension is
conjugate to arclength in the Hamiltonian, the derivative of the free energy with respect to mass gives us mean arclength that we seek.
\begin{equation}
\label{arclength-from-F}
\langle L \rangle = \frac{1}{T} \partial_{m} F.
\end{equation}
There is one complication.  The free energy is divergent in the limit that $k \rightarrow 0$.  We require the arclength of the unconfined
filament in order to compute $\Delta L$, as defined in Eq.~\ref{definition-excess-length}.  To address this problem, we compute the mean
arclength at finite $k$, take the necessary derivative, and then take the limit $k \rightarrow 0$, which then provides a finite result.

We first recall the (possibly complex) frequencies $\omega_{1,2}$ and use them to
compute derivatives of the free energy in Eq.~\ref{MHO-free-energy}.
\begin{equation}
\frac{\partial_m F_{M H O}}{T}  = \frac{\partial_m  \omega_1 L_0  \cosh (\omega_1 L_0 /2)}{2 \sinh (\omega_1 L_0 /2)} +\frac{\partial_m  \omega_2 L_0  \cosh (\omega_2 L_0 /2)}{2 \sinh (\omega_2 L_0 /2)}
\end{equation}
We know that $\omega_1^2 + \omega_2^2 = \frac{m}{\ell_{\rm P}} $ and that $\omega_1 \omega_2 = \sqrt{\frac{k}{\ell_{\rm P}}} $. When there
is  no confining potential $k=0$, one frequency vanishes. This implies that the spectrum of the
corresponding operator includes a zero eigenvalue, which, as discussed above, will be problematic for the analysis.
In preparation for taking the $k \rightarrow 0$ limit  we keep only the lowest terms in $k$ here.
Expanding to lowest order in $k$ we find that
\begin{equation}
\omega_1^2 + \omega_2^2  \pm 2 \omega_1 \omega_2  = \frac{m}{\ell_{\rm P}} \pm 2 \sqrt{\frac{k}{\ell_{\rm P}}},
\end{equation}
which means
\begin{equation}
\omega_1 \pm \omega_2   = \sqrt{ \frac{m}{\ell_{\rm P}} \pm 2 \sqrt{\frac{k}{\ell_{\rm P}}}}
\end{equation}
Solving for $\omega_{1,2}$ we find that
\begin{eqnarray}
\omega_1 &=& \sqrt{ \frac{m}{\ell_{\rm P}}} \\
 \omega_2   &=& \sqrt{ \frac{k}{m}}.
\end{eqnarray}
These results allow us to take the appropriate $m$ derivatives:
\begin{eqnarray}
\partial_m  \omega_1  &=& \frac{1}{2} \sqrt{ \frac{1}{m \ell_{\rm P}}}\\
\partial_m  \omega_2   &=& - \frac{1}{2} \sqrt{ \frac{k}{m^3 }}.
\end{eqnarray}
By assuming that the filament is long we may take $\omega_1 L_{0} \gg 1 $. We also assume that $k$ may be sufficiently small so that
$\omega_2 L _{0}\ll 1 $ (we will take it to zero shortly). The free energy of the filament is then
\begin{equation}
F_{M H O }  = T \omega_1 L_{0} /2  +T  \ln \left( \omega_2 L_{0}  \right).
\end{equation}
Taking the derivative as shown in Eq.~\ref{arclength-from-F}, we obtain the mean length
\begin{equation}
\label{arclength-k=0}
\left. \langle L \rangle \right|_{k=0}  =  \frac{1}{4} \sqrt{ \frac{1}{m \ell_{\rm P}}}  L_{0}   - \frac{1}{2 m}.
\end{equation}
For large $L_{0}$ the last term can be ignored, leaving us with the $k=0$ result.

We now turn to the case where $k$ remains finite. Then for long filaments we have $\omega_1 L  \gg 1$ and  $\omega_2  L \gg 1$,
allowing us to write $F_{MHO}$, defined in Eq.~\ref{MHO-free-energy} as
 \begin{equation}
F_{M H O }  = T (\omega_1 + \omega_2) \frac{L_0}{2} =    \frac{T L_0}{2}  \sqrt{\frac{m}{\ell_{\rm P}} + 2 \sqrt{\frac{k}{\ell_{\rm P}}}}.
\end{equation}

Then, taking the $m$ derivative as above, we obtain the mean arclength of the filament with a confining potential $k \neq 0$:
\begin{equation}
\langle  L\rangle_{M H O }  = \frac{1}{\sqrt{m \ell_{\rm P}}} \frac{1}{ \sqrt{1 + 2 \phi }  }   \frac{L_0}{4}
\end{equation}
Subtracting the equivalent quantity for the $k=0$ case, shown in Eq.~\ref{arclength-k=0}, we obtain an expression for the change in
excess mean arclength due to the presence of the confining potential
\begin{equation}
\Delta L_{M H O }  = \frac{1}{\sqrt{m b}} \frac{1}{ \sqrt{1 + 2 \phi }  }   \frac{L_0}{4} - \frac{1}{4} \sqrt{ \frac{1}{m \ell_{\rm P}}}  L_{0}
\end{equation}
Returning the original, physical parameters of the semiflexible filament model, this expression becomes
\begin{equation}
\Delta L_{M H O }  = \frac{1}{\beta \sqrt{\tau \kappa}}\left( \frac{1}{ \sqrt{1 + 2 \phi }  }  - 1\right) \frac{L_0}{4}.
\end{equation}

\section{Replica Trick}
\label{app:replica}
The core idea of the replica trick is the following. We want to find  $\left[ \ln Z \right] $ where the average is over the
ensemble of quenched potentials.  We may rewrite this in the following form
\begin{equation}
\label{basic-replica-trick}
\left[ \ln Z \right] =
\lim_{R \rightarrow 0} \left[  \frac{Z^R - 1}{R} \right]
\end{equation}
In many cases the straightforward calculation of the left hand side is intractable, while it is possible to compute on the right hand side for positive
integer  $R$ and then take the limit. The reader is referred to Ref.~\cite{Mezard:87} for further details of this approach.  We
first demonstrate the utility of the replica method in a simplified model.  We assume a Gaussian Hamiltonian so that the
partition function has the form
\begin{equation}
Z = \int {\cal D} y(x) e^{-\int y A y + 2 y_0 B y }
\end{equation}
where $A$,$B$ are operators acting on the variable $y$. The variable $y$ is affected by a quenched potential, $y_0$ whose statistics are
fixed by the Gaussian distribution
\begin{equation}
{\cal P}(V) \propto   e^{-\int y_0 C y_0 },
\end{equation}
controlled by another operator $C$.    This problem is designed to be sufficiently simple that it can be solved without the replica trick. It is
also directly related to valley approximation, as the reader may confirm.   A straightforward
calculation of the partition function, followed by taking the logarithm and then averaging the resulting free energy
over the quenched potential distribution leads to
\begin{equation}
\label{model-exact}
\left[ \ln Z \right] = \ln{ \frac{1}{\sqrt{\det A}} } - \frac{1}{2} \mbox{tr}  (C^{-1} B A^{-1} B ).
\end{equation}

If we now repeat the calculation using the replica trick, we first replicate the Hamiltonian to form $Z^{R}$:
\begin{equation}
Z^R = \int {\cal D} y_r(x) e^{-\int \sum_{r=1}^R y_r A y_r +2  y_0 B y_r },
\end{equation}
where the replicated variables are indexed by $r$: $y_r, \, r = 1,\ldots, R$.
We now average the replicated partition sum over the quenched potential (notice we are
swapping the order of the two averages) to  obtain
\begin{equation}
\label{replicated-Z}
\left[ Z^R \right] = \frac{\int {\cal D} y_r(x) {\cal D} y_0(x) e^{-\int \sum_{r=1}^R y_r A y_r + 2 y_0 B y_r + y_0 C y_0} }{ \int D y_0(x) e^{-\int y_0 C y_0 } }.
\end{equation}

Introducing $\vec{y} = \left( y_0 ,y_1 ... y_R   \right)$ we can write Eq.~\ref{replicated-Z} in the form
\begin{equation}
\left[Z^R\right] = \frac{\int {\cal D} \vec{y}(x) e^{-\int_0^T \vec{y} \mathcal{A} \vec{y} } }{\int D y_0(x) e^{-\int y_0 C y_0 } }
\end{equation}
where the matrix $\mathcal{A} $ can be written in the block form
\begin{equation}
\mathcal{A}  =
\begin{pmatrix}
C & \tilde{B}^T \\
\tilde{B} & \tilde{A}
\end{pmatrix}.
\end{equation}
The $\tilde{A}$ block is the $R \times R$ matrix coupling the replicated variables.  The $\tilde{B}$ block is an $R \times 1$ matrix
coupling the $R$ replicated variables to the quenched disorder $y_0$. The remaining calculation involves performing the Gaussian
integrals over $\vec{y}$.  We find
\begin{equation}
[Z^R] =\left( \frac{\det \mathcal{A} }{\det C } \right)^{-1/2}
\end{equation}
The determinant of the block matrix can be written as
\begin{equation}
\det \mathcal{A} = \det{\tilde{A}} \det \left(C -  \tilde{B}^T \tilde{A}^{-1} \tilde{B} \right),
\end{equation}
and we observe that, in our model, the variables $y_r$, $r = 1,\ldots, R$ are noninteracting so that the $\det{\tilde{A}} =\det A^{R}$.
As a consequence,
 \begin{equation}
[Z^R] =\left\{ \det{A}^R \det(1 - R B^{T} A^{-1} B C^{-1} ) \right\}^{-1/2}.
\end{equation}
By taking the $R \rightarrow 0$, we reproduce the exact result, Eq.~\ref{model-exact}.

Now, we consider using the replica trick for the problem at hand. Specifically, we examine a
flexible polymer at temperature $T$ interacting with a quenched
delta-correlated random potential. The polymer has tension $\tau$ and length $L$.
We will examine the average potential energy of the filament in the pinning potential.
For later computational  convenience, we introduce a coupling constant $\alpha$ controlling the interaction of the
polymer with pinning potential.

The partition function is
\begin{equation}
Z_\alpha = \int {\cal D} y(x)  e^{- \beta \int\limits_0^L  dx \frac{\tau}{2} \dot{y}^2 + \alpha  V(x,y) }
\end{equation}
We can calculate the average potential energy as
\begin{equation}
\left. \frac{\partial F }{\partial \alpha}\right|_{\alpha = 1}  = \left\langle \int V(x,y(x)) dx \right\rangle.
\end{equation}
After this thermal average, we take the average of the potential over the ensemble of pinning potentials
\begin{equation}
\label{def-average-potential}
\left. \frac{\partial [F]}{\partial \alpha}\right|_{\alpha = 1}  = \left[ \left\langle \int V(x,y(x)) dx \right\rangle \right]
\end{equation}
We assume that averaging and differentiation commute.
We expect that the mean potential energy should be negative. Indeed, the random potential will pick
negative and positive values with the same probability. The polymer, however,  will prefer the negative regions to the positive.

We now perform the necessary averages using the replica trick with, as before, the replica index being $r=1..R$. The replicated
partition sum is now
\begin{equation}
Z_\alpha^R = \int {\cal D} y_i(x)  e^{- \beta \int\limits_0^L  dx \sum\limits_{i=1}^R \frac{\tau}{2} \dot{y}^2_r + \alpha  V(x, y_r(x)) }.
\end{equation}
The average over the pinning potential, once again, couples the replicas.  We first write that average over the distribution
of $V(x,y)$,
\begin{widetext}
\begin{equation}
\left[ Z_\alpha^R \right]  = \int {\cal D} y_{r}  e^{- \beta \frac{\tau}{2}  \int_0^L  dx \sum\limits_{i=1}^R  \dot{y}_{r}^{2}}
\int {\cal D} V {\cal P} (V)  e^{- \beta \int_0^L  dx \int dy \sum\limits_{i=1}^R \alpha  V(x ,y) \delta(y -y_r(x)) }
\end{equation}
\end{widetext}
Since ${\cal P}(V)$ is a Gaussian with fixed variance $\sigma$ we can directly perform the Gaussian integral over $V$.  We
also notice that the delta functions $\delta(y - y_{r}(x))$ allow us to do the integral over $y$ in the exponent.  Doing both steps we
obtain
\begin{equation}
\label{Z-replicated-averaged}
\left[ Z_\alpha^R \right] = \int {\cal D} y_r   e^{ -\beta (\int\limits_0^L  dx \sum\limits_{i=1}^R \frac{\tau}{2} \dot{y}_{r}^2  -
\sigma \beta \alpha^2 \sum\limits_{i,j=1}^R      \delta(y_i(x) -y_j(x))   }.
\end{equation}

Eq.~\ref{Z-replicated-averaged} now appears to be the partition function of $R$ particles
 interacting through a delta-potential. It can be written in the form
\begin{equation}
\label{partition-function-Kardar}
\left[ Z_\alpha^n \right] = \int {\cal D} y_r   e^{ - L_{0} H\left(y_{1}(x), \ldots, y_{R}(x) \right)  }
\end{equation}
where the multi-particle Hamiltonian is
\begin{equation}
H = \sum_{r=1}^R  \frac{1}{2 \beta \tau} p_r^2  -  \sigma \beta^2 \sum_{i,j =1}^R   \alpha^2   \delta(y_j(x) -y_i(x)).
\end{equation}
Notice that in Eq.~\ref{partition-function-Kardar} that the length of the polymer $L_{0}$ plays the role of inverse temperature $1/T$
for the fictitious particles. Of course, $L_{0} \neq \beta$, which is the inverse temperature of the polymer.  We also note that
$x$ plays the role of time in this dynamical system of interacting particles.

Since the sum contains the term $i=j$, there is a positive infinite constant $C$ in the Hamiltonian. Separating these terms
explicitly we rewrite the Hamiltonian as
\begin{equation}
\label{repham}
H = \sum_{i=1}^R \frac{1}{2 \beta \tau} p_i^2 -  2 \sigma \beta^2 \sum_{i < j } \tau^2  \alpha^2   \delta(y_j(x) -y_i(x)) -
C \sigma \beta^2 \tau^2  \alpha^2.
\end{equation}

Reflecting on the fact that the path integral in Eq.~\ref{partition-function-Kardar} is analogous to the quantum transition amplitude
for the system of fictitious particles ~\cite{Polyakov:xx} and that the long polymer limit $L_{0} \rightarrow \infty$
corresponds to the zero temperature limit for that system, we may focus on the ground state wavefunction \cite{Kardar:87,Kardar:07}.
To find the ground state wavefunction, we use  the Bethe ansatz:
\begin{equation}
\label{wavefunction}
\Psi_0 = C_0 \exp \left( - \kappa \sum_{\alpha < \beta } | y_\alpha - y_\beta | \right).
\end{equation}
The constant $C_{0}$ is chosen to
satisfy the normalization condition. The constant $\kappa$ is set so that the discontinuity of the first derivative
cancels the delta-function, i.e., $ \frac{2}{ \beta \tau} \kappa =2  \sigma \beta^2  \alpha^2  $ .

We transform the sum by choosing a particular ordering of the $R$ particles on the $y$ line: $y_{1}<y_{2}<\ldots<y_{R}$.  In this
ordering the sum in Eq.~\ref{wavefunction} is particularly simple:
\begin{equation}
\sum_{\alpha < \beta } | y_\alpha - y_\beta | = \sum_\alpha (2 \alpha - n -1  ) y_\alpha.
\end{equation}
From this we obtain the ground state energy
\begin{equation}
E_R =  R C \sigma \beta^2 \tau^2  \alpha^2   - \frac{1}{2 \beta \tau}  \kappa^2 \frac{R (R+1) (R-1)}{3}
\end{equation}

Then the partition function is
\begin{equation}
Z^R = e^{-\beta E_R}
\end{equation}
and the free energy is

\begin{equation}
\beta F = -[ \log Z] = -\lim_{R \rightarrow 0} [ \frac{Z^R - 1}{n} ] =  \beta \lim_{R \rightarrow 0} \frac{E_R}{R}.
\end{equation}
Taking the limit we get
\begin{equation}
F   = C \sigma \beta \tau^2  \alpha^2  + \frac{1}{6}  \sigma^2 \beta^4  \alpha^4 \tau.
\end{equation}
Using Eq.~\ref{def-average-potential}, we obtain the potential energy of the thermalized polymer in the random
pinning potential, averaged over an ensemble of such potentials:
\begin{equation}
\left[ \left\langle \int V(x,y(x)) dx \right\rangle \right]   =  2 C \sigma \beta \tau^2   + \frac{2}{3}  \sigma^2 \beta^4 \tau
\end{equation}
This potential energy is positive definite, which is clearly unphysical. The result shows that the replica trick fails to find the
averaged potential energy of the polymer in the pinning potential.
The problem may lie in either the introduction of the infinite constant $C$, or in the replica trick itself.

We conjecture that the failure of the replica trick in this case is related to the failure of the analytical
continuation inherent in the method.  To expand on
this idea we consider an example related to directed polymers (without a bending energy) in a complex random potential with zero correlation
length ($\xi \rightarrow 0$), as explored by Zhang~\cite{Zhang:89}.  In that work [Eq.~(11) of that article] the author asserts that
\begin{equation}
\label{Zhang-integral}
\int dF P(F) e^{-n F} = e^{-t(n F_0 - a n^3)},
\end{equation}
where $P(F)$ is the probability distribution of free energies of the directed polymer due to the statistical ensemble of quenched potentials in its
environment and $n$ is the replica index.   The left hand side of this expression is $\left[ Z^{n} \right]$.
Now $P(F)$ should decay exponentially or faster for negative $F$, otherwise the integral will be divergent.
The result on the right hand side, arrived at through a saddle point
evaluation of the integral, is exact for all positive integer $n$ in the limit of large $t$. The goal (as in our case too) is to
examine the $n\rightarrow 0$ limit.

To examine this point let us perform the integration in Eq.~\ref{Zhang-integral} by first Taylor expanding the exponential and then integrating
term by term.
\begin{equation}
\int dF P(F) \sum_{k=0}^{\infty} \frac{ (- n F)^k}{k!}  =  \sum_{k=0}^{\infty} \frac{ (-(n F_0 - n^\beta))^k}{k!} .
\end{equation}
From Ref.~\cite{Kolokolov} we know that the large positive $F$ behavior of $P(F)$ is  $\sim \exp(-F^{\frac{5}{2}})$, ensuring the convergence of these
integral and justifying the swapping of the order of summation and integration.  The sum on the right hand side can be reorganized in terms of powers
of the replica index $n$ so that each term in the summand takes the form $d_{k}n^{k}$.  In this form that sum
can be combined with integral on the left to write an infinite sum that is now equal to zero.
\begin{equation}
\label{zero}
\sum_{k=0}^{\infty} c_k n^k = 0
\end{equation}
where
\begin{equation}
c_{k} = \int dF P(F)\frac{ \left( -F\right)^{k}}{k!} - d_{k}.
\end{equation}

If we were allowed to assert that each term in the sum Eq.~\ref{zero} separately vanished, then we would be able to calculate all moments
of the free energy directly from the known coefficients $d_{k}$.  This process for $k=1$ is the essence of the replica trick as used to compute
the mean free energy.  Unfortunately, this conclusion is not necessarily valid.  A simple counter example can be obtain from the Taylor expansion
of the function $F(z) = f(z) \sin( \pi z) $, where $f(z)$ is an analytic function.  Let $a_k$ be Taylor expansion coefficients for $F(z)$. Then it is easy to
see that if we choose $c_k = \frac{a_k}{k!} $ we will get Eq.~\ref{zero}, but not all $c_k$ are zero. The same issue appears in Ref.~\cite{Kardar:87}.

\bibliography{bib_kei}

\end{document}